\documentclass[12pt]{iopart}
\expandafter\let\csname equation*\endcsname\relax
\expandafter\let\csname endequation*\endcsname\relax
\usepackage{iopams,graphicx,braket,mathtools,array}
\usepackage{float}
\usepackage{multirow}
\usepackage[]{units}
\usepackage{hyperref}
\hypersetup{
    colorlinks=true,
    linkcolor=blue,
    filecolor=blue,      
    urlcolor=blue,
    citecolor=blue
}
\usepackage[font=small,
   justification=Justified,
   format=plain]{caption}
\usepackage{dcolumn}
\usepackage{bm}
\usepackage[dvipsnames]{xcolor}
\usepackage[caption=false]{subfig}
\usepackage{mathrsfs}
\usepackage{bbold}
\usepackage{cite}

\begin{document}
\title{Universal dissipators for driven open
quantum systems and the correction to linear response}
\author{Lorenzo Bernazzani, Bal\'azs Gul\'acsi and Guido Burkard}
\address{Department of Physics, University of Konstanz, D-78457 Konstanz, Germany.}
\ead{lorenzo.bernazzani@uni-konstanz.de (corresponding author),\\
balazs.gulacsi@uni-konstanz.de, guido.burkard@uni-konstanz.de}
\vspace{10pt}
\begin{indented}
\item[]\today
\end{indented}


\begin{abstract}
We investigate in parallel two common pictures used to describe quantum systems interacting with their surrounding environment, i.e., the stochastic Hamiltonian description, where the environment is implicitly included in the fluctuating internal parameters of the system, and the explicit inclusion of the environment via the time-convolutionless projection operator method. Utilizing these two different frameworks, we show that the dissipator characterizing the dynamics of the reduced system, determined up to second order in the noise strength or bath-system coupling, is composed of two parts. One is universal, meaning that it keeps the same form regardless of the drive term. This form constitutes the relevant part of the dissipator only as long as the drive is weak. We thoroughly discuss the assumptions on which this treatment is based and its limitations. Then, by considering the first non-vanishing higher-order term in our expansion, we derive the other, drive-dependent, term completing the full dissipator. This part of the dissipator, originating from the third cumulant, is usually neglected when modeling the decoherent dynamics of controlled qubits. However, this further term constitutes the linear response correction due to memory-mediated environmental effects in driven-dissipative quantum systems. Also, it notably shows that the structure of our quantum master equation goes beyond the Lindblad form. The Lindblad form is recovered for memory-less baths. Finally, we demonstrate this technique to be highly accurate for the problems of dephasing in a driven qubit and for the theory of pseudo-modes for quantum environments.
\end{abstract}

\maketitle

\section{Introduction}
Decoherence is the classical \emph{némesis} of quantum information systems. Nonetheless, the microscopic quantum systems utilized for quantum information processing are unavoidably immersed in an environment \cite{Schlosshauer2019}, and therefore they are coupled to some uncontrollable external degrees of freedom \cite{BurkardRMP2023}. This interaction is responsible for the loss of coherence in quantum systems \cite{Burkard2008,Schlosshauer2019}. In addition to efforts to isolate microscopic systems from unwanted perturbations, accurately modeling the effects of the environment on the evolution of the reduced system is helpful not only for mitigating these side effects, but also for enabling reservoir-engineering schemes \cite{IgnacioCirac2009,Devoret2013,Kapit2017,Murch2022}. Apart from these pragmatic aspects, the study of open quantum systems has its own fame as a fundamental problem of formidable complexity. Its complexity stems from the details of the interaction with the environment, which gives rise, e.g., to non-Markovianity \cite{Rivas_2014,BreuerRMP2016}, or from the many-body nature of the physical system, which leads to computational hardness due to the high dimension of the system's Hilbert space and to the bath model \cite{Guerreschi2020,Zagoskin2021}. As a consequence, the quantum simulation of dissipation has sparked much interest \cite{Lloyd96,Daley2014,NoriRMP2014,DelCampo2017,NoriNPJ2018,Daley2022,Jordan2022,Kockum2025}. A widely popular approach involves adding classical noise to an analog system \cite{DelCampo2017,Chenu2024}, which can yield faithful descriptions of the decoherent dynamics of the reduced system.
More generally, a quantum mechanical model of the environment may be essential to give a more complete description of the dissipative dynamics \cite{NoriNPJ2018,Jordan2022,Burkard2023,Petta2024}, thus extending the open-system theory to encompass what is generally referred to as quantum noise.

In their classic papers Gorini, Kossakowski, Sudarshan \cite{Gorini76}, and Lindblad \cite{Lindblad76} laid the rigorous mathematical foundations for treating dissipative processes in Markovian open quantum systems. However, in this treatment, the system Hamiltonian is required to be time-independent.
In most cases of interest, however, open-system dynamics occurs in the simultaneous presence of a coherent drive. In this paper, we address the question of how the drive interferes with the description of the dynamics of open systems. Notably, we address the problem in either of its aforementioned variations, i.e., with classical and quantum noise. The study of the interplay between coherent drives and random fluctuations has already a long history and has been amenable to important applications in the development of quantum information systems and quantum technologies, such as dynamical decoupling \cite{Lloyd99,Zanardi99,Vitali99,Lidar2005,DeLange2010} and quantum stochastic resonance \cite{Coppersmith94,Hanggi98,Plenio2007,Haug2019}.

The first inquiries on the dynamics of driven-dissipative systems may be traced back to the field of magnetic resonance in fluid samples \cite{Kubo54,Bloch57,Redfield57}, where in particular the interplay of the coherent radio-frequency drives and the perturbation caused by the random motion of the surrounding fluid environment was modeled \cite{AbragamBK}. The general problem was treated in a more systematic way starting with the seminal papers in Refs. \cite{Mukamel78} and \cite{Kosloff95}. In particular, in \cite{Kosloff95} the problem of a spin 1/2 system interacting with both a general bath and a strong electromagnetic field is treated, deriving the master equation for such a system. Their procedure involves the following steps: transformation to a rotating frame and subsequent diagonalization of the effective spin Hamiltonian; transformation to a further interaction picture with respect to the free-bath Hamiltonian; perturbative expansion in the bath-system coupling constant; tracing out of the bath degrees of freedom (DoF); and finally, application of the Markov approximation along with the neglect of fast-rotating terms. Overall, this procedure leads to a dissipator that contains time-dependent terms resulting from the reversion to the Schr\"odinger picture. This is also discussed in detail in a more recent work \cite{Gulacsi2024}. In the case of weak driving, this procedure is often modified, as studied, e.g., in Refs.~\cite{Goldstein2019,ClerkQuantum2023}. In fact, the standard approach that has emerged in the community is to work entirely in the laboratory frame, make use of the undriven system Hamiltonian when deriving the master equation in the interaction picture, insert the drive into the von Neumann part, and discard the fast-rotating components, thereby applying the secular approximation (SA) \cite{Goldstein2019}. This strategy closely mimics the conventional treatment of similar problems in magnetic resonance with radio-frequency irradiation \cite{SlichterBK}. An alternative approach is transforming into the frame rotating with the drive, then diagonalizing the system Hamiltonian and applying the rotating wave approximation (RWA) \cite{Goldstein2019}. When a driven system is coupled to an environment, the RWA is treated as insensitive to this frame passage, and the time-dependent factors arising from the coordinate transformation are usually not taken into account in the common analysis \cite{Goldstein2019,ClerkQuantum2023}. In fact, the noise realization is effectively defined or inserted \emph{a posteriori} in the rotating frame \cite{ClerkQuantum2023}. However, since one always perturbs the system in the lab frame, these time-dependent factors should be taken into account. Moreover, these SA and RWA approaches may lead, for the reasons just exposed, to an incorrect modeling in the case of correlated noise acting along different directions. In this work, we demonstrate how these issues can be resolved in a simple way through the formalism we develop below.
\begin{figure}[t]
\centering
\includegraphics[width=0.7\textwidth]{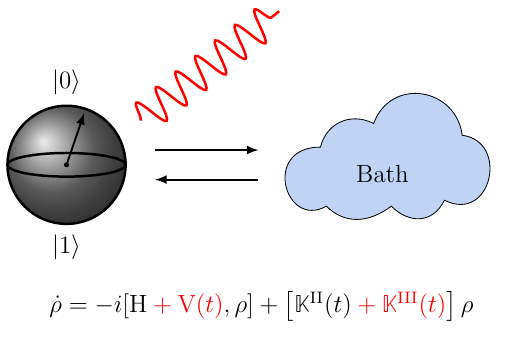}
\caption{Schematic representation of the driven open quantum system model. For ease of visualization, we depict a qubit (i.e., the system) interacting with a bath while undergoing a coherent operation via the red drive-signal. The time-local master equation corresponding to this model is displayed at the bottom. In black, we show the standard quantum-optical master equation for open systems without any time dependence in the system's Hamiltonian \cite{PetruccioneBK}. In red, we highlight the additional terms that arise due to the simultaneous presence of a coherent drive acting on the system's DoFs. In particular, the drive term is added to the bare Hamiltonian of the isolated system in the von Neumann part of the equation, as it was anticipated by previous literature, but only by heuristic arguments. On top of this, the standard dissipator, $\mathbb{K}^{\mathrm{II}}(t)$, is modified by the term $\mathbb{K}^{\mathrm{III}}(t)$. We refer to this additional term as the third-order dissipator, second-order in the dissipation + first-order in the drive.}
\label{cartoon}
\end{figure}

Our analysis shows that this procedure can indeed be notably simplified due to the combination of the coherent drive unitary with the bath-system coupling operators in the dissipator, along with the properties endowed by the null average of the effective noise process induced by the bath. This occurs without requiring \emph{a priori} knowledge of the unitary coherent dynamics of the system. In fact, our derivation is completely agnostic of what would be the unitary dynamics in the absence of noise. Therefore, it is distinct from approaches that rely on, e.g., the Magnus expansion of the unitary in the dissipator, such as those presented in Refs.~\cite{Kosloff95,Geva2004,Goldstein2019,ClerkQuantum2023,Gulacsi2024}.
We will show that our approach ultimately leads to a modification of Kubo's linear response theory \cite{Kubo66} for driven quantum systems, or equivalently in this context, to modifications to the Bloch-Redfield theory of relaxation \cite{Redfield65}. As we discuss in this work, this extends the standard linear response theory to its fully non-Markovian form by adding a term that accounts for the fluctuations that enter the description of the reduced dynamics due to the drive, thus modifying the dissipator. The result of this is the modification of the master equation for open systems, graphically summarized in Fig.~\ref{cartoon}. The addition of a coherent drive to the dissipative dynamics of the system leads to a modification of the structure of the time-local master equation consisting of adding the drive term in the von Neumann part of the equation to the bare Hamiltonian of the isolated system and adding a term $\mathbb{K}^{\mathrm{III}}$, modifying the common dissipator $\mathbb{K}^{\mathrm{II}}$. This additional term is second-order in the bath-system coupling (or noise strength), making it rightfully part of the full dissipator of the open system, and first-order in the drive strength, making it relevant for the description of the linear response of the system. We refer to it as a third-order dissipator, as it originates from the third-order cumulant of the stochastic process. It combines second-order terms in the dissipation strength with first-order terms in the drive strength, giving rise to a correction to the linear response theory of driven open systems.

The structure of the paper is as follows: first, we introduce the analytical results of our paper in Section \ref{sec:analytics} both in the stochastic Hamiltonian case and in the projection-operator formalism or Nakajima-Zwanzig description\cite{ZwanzigBK} of the reduced system dynamics in presence of a bath with quantum mechanical DoFs. The stochastic Hamiltonian case is treated in Section \ref{sec:noisy}, where we derive our master equation considering a quantum system subject to noisy parameters, i.e., classical noise. In Section \ref{sec:bath}, we derive the master equation considering a quantum system coupled to a general environment using the projection-operator technique. Then we present different examples where these results can be showcased. In particular: in Section \ref{sec:qubitdeph}, we apply our open system theory to model the dissipative dynamics of a driven qubit with dephasing induced by fluctuating parameters in the longitudinal direction and in Section \ref{sec:qubitrelax} to model the dissipative dynamics of a driven qubit with relaxation processes induced by a bath of harmonic oscillators at zero temperature.

\section{Analytical treatment of the open system dynamics}
\label{sec:analytics}

\subsection{Stochastic Hamiltonian description --- classical noise}
\label{sec:noisy}

\subsubsection{Second order master equation. ---}

We consider a system governed by the following Hamiltonian \cite{AbragamBK,VanKampenBK}:
\begin{equation}
\mathrm{H}(t)=\mathrm{H}_0+\delta\mathrm{H}_d(t)+\eta\mathrm{H}_s(t),
\label{eqhamiltonian}
\end{equation}
where $\mathrm{H}_0$ is the undriven system Hamiltonian, $\mathrm{H}_d(t)$ is the driving Hamiltonian, and $\mathrm{H}_s(t)$ is the stochastic Hamiltonian representing the influence of the environmental noise on the system. The parameters $\delta$ and $\eta$ characterize the strength of the drive and the stochastic noise, respectively. Therefore, we will restrict to Hermitian stochastic perturbations of the system. We can construct the usual time-local equation for the evolution of the noise-averaged state, that is, the state of the reduced system. This stochastic framework can be thoroughly justified considering the stochastic generalization of the Hilbert space \cite{Petruccione99,Burgarth2017}. Now, assuming that the operators are bounded, if $\delta||\mathrm{H}_d||\sim\eta||\mathrm{H}_s||\ll ||\mathrm{H}_0||$, where $||\circ||$ is a norm of the Banach space of linear operators acting on the Hilbert space of the problem, e.g., we can use the norm for bounded operator given by the following definition: $||\mathrm{H}_i||\equiv\mathrm{sup}_{\rho\in\mathcal{H}}||\mathrm{H}_i\rho||/||\rho||$, where the norm on the RHS is just the Euclidean norm of the Hilbert space \cite{AkhiezerBK} (for the case of unbounded operators, see the following Section~\ref{sec:bath}). If these conditions hold, we can go into the interaction picture with respect to $\mathrm{H}_0$ only. We further assume $\langle\mathrm{H}_s(t)\rangle=0$. We note here that in this paper, averages denoted by $\langle\,\circ\,\rangle$ are always over the realizations of the noise, while quantum averages are always expressed as ${\rm Tr}(\,\circ\,)$. There is no loss of generality in assuming a zero average, since every nonzero average part could be absorbed into the system Hamiltonian \cite{VanKampenBK}. This is a quite general description of systems usually investigated for quantum simulation and computation purposes \cite{Lloyd96,Daley2014,DelCampo2017}. We do not assume here any particular spectrum for the noise. Due to this, our treatment is indeed quite general and can be suitable for various physical situations.

From Eq.~\eqref{eqhamiltonian} we can write the von Neumann equation for the density matrix defined above, i.e., $
\dot{\rho}(t)=-i\bigl[\mathrm{H}(t),\rho(t)\bigr]\equiv-i\mathcal{L}(t)\rho(t)$. In order to solve this system of stochastic differential equations, we rely on the cumulant expansion method. This method was developed by multiple authors in various ways, see Refs.~\cite{Fox75,Fox76,VanKampen74,Terwiel74,Fox86,Skinner87}. However, we apply it in a slightly different fashion as we show next.
To start, we switch to the interaction picture and get rid of the trivial part of the Hamiltonian; thus, we set: $\rho'(t)= \exp(i\mathrm{H}_0t)\rho(t)\exp(-i\mathrm{H}_0t)$, and $\mathrm{H}'(t)=\delta\,\mathrm{H}_d'(t)+\eta\,\mathrm{H}'_s(t)$, where $\mathrm{H}'_{d/s}(t)=\exp(i\mathrm{H}_0t)\mathrm{H}_{d/s}(t)\exp(-i\mathrm{H}_0t)$ are the interaction picture Hamiltonians for the drive and stochastic perturbations, respectively. In the interaction picture, the von Neumann equation reads:
\begin{equation}
\dot{\rho}'(t)=-i\mathcal{L}'(t)\rho'(t)=-i\delta\bigl[\mathrm{H}_d'(t),\rho'(t)\bigr]-i\eta\bigl[\mathrm{H}_s'(t),\rho'(t)\bigr] =-i\bigl[\mathcal{L}'_d(t)+\mathcal{L}'_s(t)\bigr]\rho'(t).
\label{eqliouville}
\end{equation}
Now we seek an iterative solution and write \cite{AbragamBK,SlichterBK,VanKampenBK}
\begin{equation}
\rho'(t)=\rho(0)-i\int_0^t\mathcal{L}'(t_1)\rho(0)dt_1-\int_0^t\int_0^{t_1}\mathcal{L}'(t_1)\mathcal{L}'(t_2)\rho'(t_2)dt_1dt_2.
\label{eqexpansion}
\end{equation}
Going on like this, after an infinite number of iterations, we get to 
\begin{equation}
\rho'(t)=\mathbb{Y}(t|0)\rho(0).
\label{eqmap}
\end{equation}
Taking the stochastic average of both sides, the equation above implies:
\begin{equation}
\langle\rho'(t)\rangle=\langle\mathbb{Y}(t|0)\rangle\rho(0),
\label{eqmapave}
\end{equation}
since $\rho(0)=\rho'(0)$ is not random, and where we have introduced the non-local kernel 
$\mathbb{Y}(t|0)=\mathbb{1}+\sum_{n=1}^{+\infty}(-i)^n\int\dots\int\mathcal{L}'(t_1)\dots\mathcal{L}'(t_n)dt_1\dots dt_n$ \cite{VanKampenBK,Skinner87}. The stochastic average in Eq.~\eqref{eqmapave} is over the realizations of the noise processes. Differentiating and assuming that $\langle\mathbb{Y}(t|0)\rangle$ is invertible\footnote{Regarding the invertibility of this kernel and its physical meaning in the construction of master equations for quantum states, the reader is referred to Ref.~\cite{Andersson2007}.} then leads to:
\begin{equation}
\langle\dot{\rho}'(t)\rangle=\langle\dot{\mathbb{Y}}(t|0)\rangle\rho(0)=\langle\dot{\mathbb{Y}}(t|0)\rangle\langle\mathbb{Y}(t|0)\rangle^{-1}\langle\rho'(t)\rangle,
\label{eqy}
\end{equation}
where $\mathbb{K}_c'(t)\equiv\langle\dot{\mathbb{Y}}(t|0)\rangle\langle\mathbb{Y}(t|0)\rangle^{-1}$ is a non-stochastic superoperator by construction, since it connects averaged quantities. In this way, we constructed a time-local master equation. We expand $\mathbb{K}_c'(t)$ in orders of $||\mathcal{L}'(t)||$. If, e.g., the system is described by a finite-dimensional Hilbert space, i.e., by an N-level system, then the expansion parameters are just $||\mathcal{L}'(t)||\sim\delta\sim\eta$, since $||\mathrm{H}'_{d,s}(t)||=||\mathrm{H}_{d,s}(t)||=1$, according to the definition of norm we gave before. Those are to be confronted with the level splitting, that is twice the prefactor in front of the bare Hamiltonian, usually named $\Omega$. In this N-level system case, one can vectorize the density matrix, then the superoperators become matrices in tetradic space \cite{MukamelBK}.
We then truncate this series in second order (Born approximation), which yields:
\begin{align}
\langle\dot{\rho}'(t)\rangle=&\,\bigl[\mathbb{K'}_c^{\rm{I}}(t)+\mathbb{K'}_c^{\rm{II}}(t)\bigr]\langle\rho'(t)\rangle\,,\quad \mathrm{where}\label{eqcumulant}\\
\mathbb{K'}_c^{\rm{I}}(t)=&-i\bigr{\langle}\mathcal{L}'(t)\bigr{\rangle}\,,\quad\mathrm{and}\quad\mathbb{K'}_c^{\rm{II}}(t)=-\int_0^{t}\bigr{\langle}\bigr{\langle}\mathcal{L}'(t)\mathcal{L}'(t-t')\bigr{\rangle}\bigr{\rangle}\,dt.
\end{align}
Here $\mathcal{L}'(t)=\mathcal{L}'_d(t)+\mathcal{L}'_s(t)$ and we introduced the cumulant symbol, which means $\bigr{\langle}\bigr{\langle}\mathcal{L}'(t)\mathcal{L}'(t-t')\bigr{\rangle}\bigr{\rangle}=\bigr{\langle}\mathcal{L}'(t)\mathcal{L}'(t-t')\bigr{\rangle}-\bigr{\langle}\mathcal{L}'(t)\bigr{\rangle}\bigr{\langle}\mathcal{L}'(t-t')\bigr{\rangle}$, moreover $\bigr{\langle}\bigr{\langle}\mathcal{L}'(t)\bigr{\rangle}\bigr{\rangle}=\bigr{\langle}\mathcal{L}'(t)\bigr{\rangle}$, as follows from the standard definition of cumulant averages in statistics \cite{VanKampenBK}.

Reverting to the Schr\"odinger picture, we finally obtain the master equation,
\begin{equation}
\langle\dot{\rho}(t)\rangle=\Bigl[-i\bigr{\langle}\mathcal{L}(t)\bigr{\rangle}-\int_0^{t}\bigr{\langle}\bigr{\langle}\mathcal{L}(t)\,e^{it'\mathcal{L}_0}\mathcal{L}(t-t')\bigr{\rangle}\bigr{\rangle}\,e^{-it'\mathcal{L}_0}dt'\Bigr]\langle\rho(t)\rangle.
\label{eqschroedpic1}
\end{equation}
Here $e^{\pm it'\mathcal{L}_0}$ is the exponential map of the (super)operator $\mathcal{L}_0$. More simply, in the case of an N-level system after vectorization, this is just the exponential of the matrix representation of $\mathcal L_0$. In short, in the interaction frame given by $\mathrm{H}_0$ we constructed a time-local master equation using the standard method of the cumulant expansion \cite{VanKampen74,Fox86,Skinner87}, with partial time-ordering prescription. Throughout, we work in the Liouvillean space, i.e., we used the linear operators $\mathcal{L}_i=\bigl[\mathrm{H}_i,\circ\bigr]$. In the case of a finite-dimensional Hilbert space, such as the one for N-level systems, the state can be vectorized and the Liouvillean can be written as a matrix. Otherwise the intended action of it is given by the commutator and the action of $e^{-it\mathcal{L}_0}$ on a state is given by the exponential map $e^{-it\mathcal{L}_0}\rho=e^{-i\mathrm{H}_0t}\rho e^{i\mathrm{H}_0t}$.

Now we demonstrate the first result of this paper by showing that the deterministic part of the drive cancels out in the second-order cumulant. Writing $\mathcal{L}(t)=\mathcal{L}_d(t)+\mathcal{L}_s(t)$, where $\langle\mathcal{L}_s(t)\rangle=0$, we have:
\begin{align}
\bigr{\langle}\bigr{\langle}&\mathcal{L}(t)\,e^{it'\mathcal{L}_0}\mathcal{L}(t-t')\bigr{\rangle}\bigr{\rangle}=\\
&=\mathcal{L}_d(t)\, e^{it'\mathcal{L}_0}\mathcal{L}_d(t-t')+\bigr{\langle}\mathcal{L}_s(t)\,e^{it'\mathcal{L}_0}\mathcal{L}_s(t-t')\bigr{\rangle}-\mathcal{L}_d(t)\,e^{it'\mathcal{L}_0}\mathcal{L}_d(t-t')\,,\nonumber
\label{eqtrick}
\end{align}
and the first and third terms on the RHS of the equation cancel.
This leads to the following form for the master equation:
\begin{equation}
\langle\dot{\rho}(t)\rangle=\Bigl[-i\bigr{\langle}\mathcal{L}(t)\bigr{\rangle}-\int_0^{t}\bigr{\langle}\mathcal{L}_s(t)\,e^{it'\mathcal{L}_0}\mathcal{L}_s(t-t')\bigr{\rangle}\,e^{-it'\mathcal{L}_0}dt'\Bigr]\langle\rho(t)\rangle.
\label{meqclassic}
\end{equation}
This formula shows that in the case of a noise perturbation of a Hermitian Hamiltonian of a weakly driven system, the drive does not enter in the second-order dissipator, but it appears at first order in the von Neumann term. In a different context, two of the authors of the present paper have already pointed out the special case of this result in Appendix A of Ref.~\cite{Bernazzani2023}.

This result has been utilized in the magnetic resonance community, particularly in radio-frequency spectroscopy, to complement the theory of motional narrowing of spectroscopic lines under coherent irradiation. Since in this framework, the main interest was in the asymptotic dynamics that give the relaxation times for the spin, this result has been used (but not proven) in the Markov approximation. Here we provided a full non-Markovian model and gave a formal proof of what has been sustained only by heuristic arguments. This is witnessed by the classic book by Slichter \cite{SlichterBK}. There, it is discussed that the usual treatment of the drive term when this has a small amplitude is to include it only to first order in the master equation. On the other hand, Abragam, in his book \cite{AbragamBK}, utilizes the non-viscous liquid approximation, which replaces the coherent unitary in the dissipator with an approximate form, rendering the problem analytically tractable. This latter approach is equivalent to a first-order Magnus expansion that has been applied recently in the literature on non-Markovian open quantum systems \cite{ClerkQuantum2023,Gulacsi2024}. These approaches are different from ours since they make use of the interaction picture with respect to the full,time-dependent Hamiltonian and then approximate the unitary in the dissipator with orders of the Magnus expansion. If the Magnus expansion is performed up to the first order, this procedure leads to the same result as the one exposed by Slichter.



Ultimately, such master equations are perturbative expansions in the noise strength or bath-system coupling, multiplied by a characteristic memory time of the environment. In contrast, our method defines its validity purely in terms of the relative magnitudes of the drive strength, the noise strength (or coupling), and the system's level spacing. Unlike the Magnus-based treatment, which relies on solving the driven coherent dynamics and then perturbing around these trajectories, our approach is conceptually different. It requires no prior knowledge of the driven evolution: we treat both coherent and incoherent terms perturbatively, starting from the unperturbed eigenbasis of the bare Hamiltonian. Furthermore, in the Markovian limit, our approach provides a formal foundation for the classical treatments found in earlier literature, such as those by Slichter and Abragam \cite{SlichterBK,AbragamBK}.

All this is closely related to the standard application of the RWA or SA \cite{SlichterBK}.
In summary, we have explained why the RWA remains valid up to second order: it is sufficient that the drive strength be comparable to that of the noise. These insights can be easily generalized to the reduced dynamics of systems immersed in an environment, with $\eta$ taking the role of the system-environment coupling in that case. In Section \ref{sec:bath}, we demonstrate this equivalence using the projection operator technique.

Note that, due to the perturbation having a non-zero stochastic average, we cannot be assured that the higher orders of the cumulant expansion vanish. Nonetheless, they remain of higher order in both the drive and noise strength. We investigate the third-order term in detail in the next subsection.

\subsubsection{Higher order contributions. ---}

Let us now delve more into the higher orders of the cumulant expansion. Note that if $\eta<\delta\ll\Omega$, the third-order term gives the first correction to the linear response theory, in the presence of coherent driving. This correction is of the same order as the usual second-order dissipator and can therefore lead to observable consequences. In the case $\eta\sim\delta\ll\Omega$, the third-order and higher contributions are only smaller corrections with respect to the parameters.

Here, we show that, besides changing the dissipator, the third-order term also leads to an interesting renormalization of the drive. Incidentally, it was already shown in the undriven case that if the system and noise Hamiltonians commute, the third- and fourth-order `terms in the cumulant expansion, Eq. \eqref{eqy}, vanish altogether \cite{Skinner87}. This also happens in the driven case if the noise Hamiltonian commutes with the drive and additionally, if the drive Hamiltonian commutes with either the system Hamiltonian or the noise Hamiltonian \cite{Llobet2018}. The following theory generalizes these special cases.

From Eq.~\eqref{eqy}, The inclusion of the third order term formally gives:
\begin{equation}
\langle\dot{\rho}'(t)\rangle=\bigl[\mathbb{K'}_c^{\rm{I}}(t)+\mathbb{K'}_c^{\rm{II}}(t)+\mathbb{K'}_c^{\rm{III}}(t)\bigr]\langle\rho'(t)\rangle,
\label{eqcumulantthird}
\end{equation}
where $\mathbb{K'}_c^{\mathrm{I}}=-i\langle\mathcal{L}'(t)\rangle=-i\delta\bigl[\mathrm{H}'_d(t),\circ\bigr]\,$ and $\mathbb{K'}_c^{\mathrm{II}}=-\int_0^{t}\bigl{\langle}\mathcal{L}'_s(t)\mathcal{L}'_s(t-t')\bigr{\rangle}\,dt'$, as we showed before. Then we must compute
\begin{equation}
\mathbb{K'}_c^{\rm{III}}(t)=i\int_0^{t}\!\int_0^{t_1}\bigl{\langle}\bigl{\langle}\mathcal{L}'(t)\mathcal{L}'(t_1)\mathcal{L}'(t_2)\bigr{\rangle}\bigr{\rangle}\,dt_2dt_1\,.
\end{equation}

In the interaction picture, the third-order term has the following form \cite{Fox76,Skinner87}:
\begin{align}
\langle\langle\mathcal{L}'(t)\mathcal{L}'(t_1)\mathcal{L}'(t_2)\rangle\rangle&=\langle\mathcal{L}'(t)\mathcal{L}'(t_1)\mathcal{L}'(t_2)\rangle-\langle\mathcal{L}'(t)\rangle\langle\mathcal{L}'(t_1)\mathcal{L}'(t_2)\rangle\\&-\langle\mathcal{L}'(t)\mathcal{L}'(t_1)\rangle\langle\mathcal{L}'(t_2)\rangle
-\langle\mathcal{L}'(t)\mathcal{L}'(t_2)\rangle\langle\mathcal{L}'(t_1)\rangle\nonumber\\&+
\langle\mathcal{L}'(t)\rangle\langle\mathcal{L}'(t_1)\rangle\langle\mathcal{L}'(t_2)\rangle+
\langle\mathcal{L}'(t)\rangle\langle\mathcal{L}'(t_2)\rangle\langle\mathcal{L}'(t_1)\rangle\nonumber\,.
\end{align}
Assuming the noise is Gaussian, we can simplify this third-order term through some algebra:
\begin{equation}
\mathbb{K'}_c^{\rm{III}}(t)=i\int_0^{t}\!\int_0^{t_1}\bigl{\langle}\mathcal{L'}_s(t)\bigl[\mathcal{L'}_d(t_1),\mathcal{L'}_s(t_2)\bigr]\bigr{\rangle}\,dt_2dt_1\,.
\label{eqdrivethird}
\end{equation}
This is the second main result presented in this paper. Equation~\eqref{eqdrivethird} serves as the correction to linear response theory for driven quantum systems subjected to Gaussian stochastic noise.

One can then easily transform the time-local equation back to the Schrödinger picture:
\begin{equation}
\langle\dot{\rho}(t)\rangle=\bigl[\mathbb{K}_c^{\rm{I}}(t)+\mathbb{K}_c^{\rm{II}}(t)+\mathbb{K}_c^{\rm{III}}(t)\bigr]\langle\rho(t)\rangle,
\label{eqdrivethirdsp}
\end{equation}
where $\mathbb{K}_c^{\rm{I}}=-i\langle\mathcal{L}(t)\rangle=-i\delta\bigl[\mathrm{H}_0+\mathrm{H}_d(t),\,\circ\,\bigr]\,$ and $\mathbb{K}_c^{\rm{II}}=-\int_0^{t}\bigl{\langle}\mathcal{L}_s(t)e^{it'\mathcal{L}_0}\mathcal{L}_s(t-t')\bigr{\rangle}e^{-it'\mathcal{L}_0}\,dt'$. The third-order term comes in a more easily readable form if we make the following substitutions for the integration variables: $t_1\to t-t'\,,\,t_2\to t-t'-t''$, which leads to:
\begin{align}
\mathbb{K}_c^{\rm{III}}(t)=i\int_0^{t}\!\int_0^{t-t'}&\Bigl[\bigl{\langle}\mathcal{L}_s(t)e^{it'\mathcal{L}_0}\mathcal{L}_d(t-t')e^{it''\mathcal{L}_0}\mathcal{L}_s(t-t'-t'')e^{-i(t'+t'')\mathcal{L}_0}\bigr{\rangle}\nonumber\\
-&\bigl{\langle}\mathcal{L}_s(t)e^{i(t'+t'')\mathcal{L}_0}\mathcal{L}_s(t-t'-t'')e^{-it''\mathcal{L}_0}\mathcal{L}_d(t-t')e^{-it'\mathcal{L}_0}\bigr{\rangle}\Bigr]\,dt''dt'\,.
\label{thirdexpl}
\end{align}
As is clear from this expansion, the third-order term contains the drive at first order. This indicates that the standard treatment of the Bloch-Redfield relaxation theory does not capture the complete linear response of the system, since it misses the term above. This is in essence what we mean by correcting linear response theory by our formalism. In fact, the second- and third-order terms of Eq.~\eqref{thirdexpl} fully describe how environmental effects influence the linear response of the system.

We now include the third-order correction to the dissipator in a simple example.
We assume to have a qubit subject to, e.g., noise in the longitudinal direction $\mathrm{H}_s(t)=\eta(t)\sigma_z$, with autocorrelation function $\langle\eta(0)\eta(t)\rangle=\eta^2c(t)$, in the usual transverse drive configuration $\mathrm{H}_d(t)=\delta\bigl[f(t)\sigma_++f^{*}(t)\sigma_-\bigr]$. We pass now to another vectorization of the density matrix, i.e., the one named after Bloch \cite{Bloch53,Feynman57}, by expressing the density matrix in the form $\rho(t)=\bigl[\mathbb{1}+ \boldsymbol{r}(t)\cdot\boldsymbol{\sigma}\bigr]/2$. In this case $i\mathcal{L}_0\to-\Omega L_z$, one of the generators of the algebra of spatial rotations, i.e.,
\begin{equation}
L_x=
\begin{bmatrix}
0 && 0 && 0\\
0 && 0 && -1\\
0 && 1 && 0\\
\end{bmatrix},\quad
L_y=
\begin{bmatrix}
0 && 0 && 1\\
0 && 0 && 0\\
-1 && 0 && 0\\
\end{bmatrix},\quad
L_z=
\begin{bmatrix}
0 && -1 && 0\\
1 && 0 && 0\\
0 && 0 && 0\\
\end{bmatrix}.
\label{so(3)}
\end{equation}
Now one can map the Liouvillean operators of the previous part to infinitesimal rotations of the Bloch sphere.
After a bit of algebra, we obtain for the third order (remembering that $L_z^2\neq\mathbb{1}_3$):
\begin{equation}
\mathbb{K}_c^{\rm{III}}(t)=4\eta^2\int_0^tdt'\int_0^{t-t'}dt''\,c(t'+t'')\,e^{-\Omega t'L_z}\Bigl[L_zL_d(t-t')L_z-L_z^2\,L_d(t-t')\Bigr]e^{\Omega t'L_z}.
\label{thirdordermedeph}
\end{equation}
The third-order term consists of a renormalization of the drive and an additional non-Hermitian part, given by the second term in parentheses.
Noticing that for a configuration with transverse drive, the first term in square brackets in Eq.~\eqref{thirdordermedeph} vanishes, the equation above explicitly reads:
\begin{equation}
\mathbb{K}_c^{\rm{III}}(t)=8\eta^2\delta\int_0^tdt'\int_0^{t-t'}dt''\,c(t'+t'')\,e^{-\Omega t'L_z}\begin{bmatrix}
0 && 0 && -f_I(t-t')\\
0 && 0 && f_R(t-t')\\
0 && 0 && 0\\
\end{bmatrix}e^{\Omega t'L_z},
\label{thirddephasing}
\end{equation}
where we wrote $f_R=\mathrm{Re}(f)$ and $f_I=\mathrm{Im}(f)$. Notably, the third-order term makes the Bloch matrix asymmetric, entering only the equations for the transverse components $\mathrm{r_{x/y}}$.
We are going to apply this formula in Section \ref{sec:qubitdeph}.

One could compute further orders of this expansion if a higher precision is required. If the Gaussianity assumption is used, there can be notable simplifications to these computations.  Nonetheless, it is important to notice that in the fourth order, shown in \ref{appendixiv}, there is no fourth-order term in the drive strength, because of the symmetry of this cumulant. There can be no third-order term, either. The contributions can be only of second-order in the drive strength (and second-order in the noise strength). Therefore, these contributions are of higher order than the one given in this section. For a more quantitative estimate of the errors made in truncating this generalized cumulant expansion at different orders, we refer the reader to \ref{appendixiv}.

\subsection{Bath-system description --- quantum noise}
\label{sec:bath}

\subsubsection{Second order TCL master equation. ---}

We consider now the presence of a thermal bath and compute the dynamical equations for the reduced system in the weakly driven case.  The results contained here are quite general and can be applied to different systems, with only a few modifications from the undriven case \cite{Burkard2004,Lidar2007,Burkard2023}. The addition of the bath results in an enlargement of the Hilbert space of the problem. The new degrees of freedom introduced will be treated dynamically. The dynamics of the system and bath combined can be generally described by the Hamiltonian of the form \cite{Petruccione99,VanKampenBK}:
\begin{equation}
\mathrm{H}(t)=\mathrm{H}_S(t)+\mathrm{H}_{BS}+\mathrm{H}_B
=\bigl[\mathrm{H}_0+\delta\,\mathrm{H}_D(t)\bigr]\otimes\mathbb{1}_B+\eta\,\mathrm{H}_{BS}+\mathbb{1}_S\otimes\mathrm{H}_B\,,
\label{eqbath}
\end{equation}
where $\mathrm{H}_{BS}$ denotes the bath-system interaction.
We assume that initially the state of the combined system and bath can be written as a product state, i.e., $\rho(0)=\rho_S(0)\otimes\rho_B(0)$. Moreover, we set $\bigl[\mathrm{H}_B,\rho_B(0)\bigr]=0$, and this state of the bath, which can be described for simplicity as a thermal state $\rho_B(0)=\rho_B=\exp(-\beta\mathrm{H}_B)/\mathrm{Tr}[\exp(-\beta\mathrm{H}_B)]$, which can be used to define a projection operator $\mathcal{P}\rho\equiv\mathrm{Tr}_B(\rho)\otimes\rho_B$ \cite{Lidar2007}.
After passing to the interaction picture, i.e., $\rho'=e^{i(\mathcal{L}_0+\mathcal{L}_B)t}\rho$ and $\mathcal{L}'=\mathcal{L}_D'+\mathcal{L}_{BS}'=\exp(-i\mathcal{L}_0t)(\mathcal{L}_D+\mathcal{L}_{BS})\exp(i\mathcal{L}_0t)$ ($\hbar=1$), the time-local master equation is \cite{Chaturvedi79}
\begin{equation}
\mathcal{P}\dot{\rho}'=\mathbb{K}_q'(t)\mathcal{P}\rho'(t)\,.
\label{eqshibata}
\end{equation}
Here, we discarded the solutions with $\bigl(\mathbb{1}-\mathcal{P}\bigr)\rho(0)\ne 0$, since we selected an uncorrelated initial state $\rho(0)$ (defined above). We emphasize that the superoperator $\mathbb{K}_q'(t)$ is not the same as in the previous Section~\ref{sec:noisy}, but we use a similar notation to highlight the correspondence between the two frameworks. The average in the classical section can be seen as just a particular type of projector. The only difference between the classical and the quantum noise case ($\mathbb{K}_c'(t)$ and $\mathbb{K}_q'(t)$) is that here we treat the bath in a quantum-mechanical way, and this generally leads to non-vanishing imaginary parts of the correlation functions of the bath that give rise, e.g., to detailed balance. The kernel $\mathbb{K}_q'(t)$ would reduce to $\mathbb{K}_c'(t)$ if we set the imaginary parts of the bath correlation functions to zero. A sufficient condition for this would be $\bigl[\mathrm{H}_{BS},\mathrm{B}\bigr]=0$. For this reason, we adopt a similar symbol for this dynamical generator in both the classical and quantum noise cases.

Next, we expand $\mathbb{K}_q'(t)$ in series up to second order in $\mathcal{L}'$, i.e., $\mathbb{K}_q'^{\rm{I}}(t)+\mathbb{K}_q'^{\rm{II}}(t)=-i\mathcal{P}\mathcal{L}'(t)-\int_0^t \bigl[\mathcal{P}\mathcal{L}'(t)\mathcal{L}'(t-t')-\mathcal{P}\mathcal{L}'(t)\mathcal{P}\mathcal{L}'(t-t')\bigr]dt'$. This expansion may bring forward the problem of the norms of the operators and the actual meaning of the strengths we defined before. In the case of spin baths, the statements about the norms of the operators can be translated quite straightforwardly from the analysis we gave for classical noise. Bosonic baths, i.e., baths of oscillators, deserve better clarification. In the case of oscillator baths, it is clear that the norm of $\mathrm{B}$ is infinite, therefore the definition of a coupling strength seems meaningless \cite{Lidar2020}. However, in our perturbation theory $\delta||\mathbb{1}_B\otimes\mathrm{H}_{D}(t)||$ and $\eta||\mathrm{H}_{BS}||$ have to be compared with $||\mathbb{1}_B\otimes\mathrm{H}_0||+||\mathbb{1}_S\otimes\mathrm{H}_{B}||$, for the series expansion to hold meaningful. If the reduced system is an N-level system, the system part of those operators is trivial since those are just rotations of the reduced system. The bath parts instead show more richness, since they depend on the form of the bath-system coupling Hamiltonian; using linear harmonic oscillators we have $\mathrm{H}_{B}=\sum\omega_ka_k^{\dagger}a_k$ and in general $\mathrm{H}_{BS}=\sum(a_k^{\dagger}\pm a_k)^n\otimes\sigma_i$, therefore we see that if the exponent in $\mathrm{H}_{BS}$ is 1 we have that $||\mathrm{H}_{BS}||\sim\sqrt{n}$ and $||\mathrm{H}_{B}||\sim n$, then $||\mathrm{H}_{BS}||\ll||\mathrm{H}_{B}||$ is fulfilled increasingly in the occupation number of the bath. In the case of low occupation numbers, the comparison of the coupling strengths and the level spacing becomes meaningful. If $n=2$, instead, the quantities to be compared are indeed trivially the coupling strength and the level spacing. If $n>2$, this picture holds only for baths close to the vacuum state, but this case rarely occurs in problems of interest. We analyzed the case of a reduced system composed of finite-dimensional (N-level) systems. This fact can be termed colloquially as the \emph{hierarchy of the norms}. The extension of this to the fully quantum-optical realm, where the reduced system is an oscillator (perhaps nonlinear), may also be interesting. The study of this case goes beyond the purposes of the present paper and is relegated to future work.

We can now prove results that are formally very similar to those exposed in the previous section. We manage to do so quite easily by starting from Eq.~\eqref{eqshibata} and by making some fairly plausible assumptions about the projector.
From the form of the starting Hamiltonian and the fact that the Hilbert space is a direct product, it follows that if we split $\mathcal{L}=\mathcal{L}_0+\mathcal{L}_D+\mathcal{L}_B+\mathcal{L}_{BS}=\mathcal{L}_S+\mathcal{L}_B+\mathcal{L}_{BS}$, we can safely assume $\mathcal{P}\mathcal{L}_B=\mathcal{L}_B\mathcal{P}=0$, $\mathcal{P}\mathcal{L}_S=\mathcal{L}_S\mathcal{P}$ 
and $\mathcal{P}\mathcal{L}_{BS}\mathcal{P}=0$ \cite{VanKampenBK}. This is analogous to stating that the dynamics induced by the bath have a null average in the sense of classical stochastic processes. This poses constraints on the state of the bath, that have to be in a Gaussian state.
In this way, we get exactly from Eq. \eqref{eqshibata}, by algebraic steps detailed in~\ref{sec:AppendixBath}, the following master equation in the Schr\"odinger picture:
\begin{equation}
\mathcal{P}\dot{\rho}=-i\mathcal{L}_S(t)\mathcal{P}\rho
-\int_0^t\mathcal{P}\mathcal{L}_{BS}(t)e^{i(\mathcal{L}_0+\mathcal{L}_B)t'}\mathcal{L}_{BS}(t-t')e^{-i(\mathcal{L}_0+\mathcal{L}_B)t'}dt'\mathcal{P}\rho
\label{eqschroedpic2}\,.
\end{equation}
Remarkably, the drive term has disappeared from the dissipator part when going from Eq.~\eqref{eqshibata} to Eq.~\eqref{eqschroedpic2}. In this treatment, the driving term drops exactly and not because of some approximation. The form we reach here, not containing any drive contribution in this dissipator, is a consequence of the time-local structure of the master equation. This is why we term this dissipator \emph{universal}. As we will show in the following, this is just one part of the full dissipator, the part that is independent on the drive.

Given the form above for the projector $\mathcal{P}$ onto the equilibrium subspace, we write $\mathcal{P}\rho=\mathrm{Tr}_B(\rho)\otimes\rho_B$. Furthermore, utilizing the linearity of the trace and $\mathrm{Tr}\bigl[\mathcal{L}_{BS}(t)\rho_S(t)\bigr]\otimes\rho_B=0$ and $\mathrm{Tr}\bigl[\mathcal{L}_D(t)\mathcal{L}_{BS}(t-t')\rho_S(t)\bigr]\otimes\rho_B=\mathrm{Tr}\bigl[\mathcal{L}_{BS}(t)\mathcal{L}_D(t-t')\rho_S(t)\bigr]\otimes\rho_B=0$, we obtain:
\begin{equation}
\dot{\rho}_S(t)=-i\mathcal{L}_S(t)\rho_S(t)-\int_0^t dt'\mathrm{Tr}_B\bigl[\mathcal{L}_{BS}(t)e^{i(\mathcal{L}_0+\mathcal{L}_B)t'}\mathcal{L}_{BS}(t-t')e^{-i(\mathcal{L}_0+\mathcal{L}_B)t'}
\rho_S(t)\otimes\rho_B\bigr].
\label{meqquantum}
\end{equation}
The two equations \eqref{eqschroedpic2} and \eqref{meqquantum} complete the proof of formal equivalence with Eq. \eqref{meqclassic}.

\subsubsection{Third order term. ---}

Let us now consider the higher orders of the expansion. Clearly, these are of a higher order with respect to the parameters $\delta\sim\eta$. Notwithstanding that a more accurate approximation could require them, we show here that, in particular, the third-order term extends linear response theory in a manner analogous to the stochastic case discussed in Section~\ref{sec:noisy}.

From Eq.~\eqref{eqshibata}, the inclusion of the third-order term formally gives
\begin{equation}
\mathcal{P}\dot{\rho}'(t)=\bigl[\mathbb{K'}_q^{\rm{I}}(t)+\mathbb{K'}_q^{\rm{II}}(t)+\mathbb{K'}_q^{\rm{III}}(t)\bigr]\mathcal{P}\rho'(t)\,,
\label{eqbaththird}
\end{equation}
where $\mathbb{K'}_q^{\rm{I}}=-i\mathcal{L}_D'(t)$ and $\mathbb{K'}_q^{\rm{II}}=-\int_0^{t}\mathcal{P}\mathcal{L}'_{BS}(t)\mathcal{L}'_{BS}(t-t')\,dt'$ as we showed before. In the interaction picture, the last term has the following form:
\begin{equation}
\mathbb{K'}_q^{\rm{III}}(t)=i\int_0^{t}\!\int_0^{t_1}\mathcal{P}\mathcal{L'}_{BS}(t)\bigl[\mathcal{L'}_D(t_1),\mathcal{L'}_{BS}(t_2)\bigr]\mathcal{P}\,dt_2dt_1\,.
\label{eqbathdrivethird}
\end{equation}
As is evident, this term is not universal in the sense that it depends on the form of the drive. As noted above, this term is fundamental when accounting for the linear response of the system. Thus, our formalism provides this missing part of the linear response of the system that has been neglected so far in the literature.

\section{Qubit with parametric dephasing}
\label{sec:qubitdeph}

As a concrete example, we consider the dissipative dynamics of a qubit, during monochromatic drive, subject to dephasing noise with Ornstein-Uhlenbeck (OU) spectrum \cite{Uhlenbeck30}. We show that within this scenario we can derive analytical equations that corroborate our more general, and system-independent, treatment given above.

The Hamiltonian of a driven two-level system (TLS) with Gaussian dephasing noise can be written simply as:
\begin{equation}
\mathrm{H}(t)=\frac{\Omega}{2}\sigma_z+\frac{D}{2}\Bigl[e^{-i(\omega t +\varphi)}\sigma_++e^{i(\omega t +\varphi)}\sigma_-\Bigr]+\eta(t)\sigma_z\,,
\end{equation}
where $\eta$ is characterized by the following two moments
\begin{equation}
\langle\eta(t)\rangle=0\,,\quad\langle\eta(t)\eta(t')\rangle=\eta^2c(t-t')=\frac{g}{4\tau}\exp\Bigl(-\frac{|t-t'|}{\tau}\Bigr).
\end{equation}
In a frame rotating along with the coherent drive, the Hamiltonian becomes:
\begin{equation}
\mathrm{H}'(t)=\frac{1}{2}\bigl[\Delta\sigma_z+D\cos(\varphi)\sigma_x+D\sin(\varphi)\sigma_y\bigr]+\eta(t)\sigma_z.
\end{equation}
As explained in Section~\ref{sec:noisy}, we can always assume that the noise is represented by a stationary stochastic process. In this way, we can apply Novikov's theorem (NT) and write a time-non-local exact master equation \cite{Novikov65,Budini2000,Budini2001,Kiely2021}:
\begin{equation}
\langle\dot{\rho}'\rangle=-i\Bigl[\langle\mathrm{H}'\rangle,\langle\rho'\rangle\Bigr]-\eta^2\int_0^{t}dt'c(t-t')\bigl[\sigma_z,\bigl[\sigma_z,\langle\rho'(t')\rangle\bigr]\bigr].
\label{eqnovikov}
\end{equation}
We use this exact form, which can be integrated analytically thanks to the special symmetry properties of this example, to show the validity and accuracy of our time-local master equation.

We rewrite Eq.~\eqref{eqnovikov} in the Bloch vector basis, $\boldsymbol{\mathrm{r}}(t)={\rm Tr}[\boldsymbol{\sigma}\rho(t)]$, and solve it by Laplace transformation, 
$\boldsymbol{\mathrm{R}}(s)=\int_0^\infty \boldsymbol{\mathrm{r}}(t)e^{-st}\,dt$,
to benchmark Eq.~\eqref{meqclassic}.
Laplace-transforming the differential equation above, we get the following algebraic equation:
\begin{equation}
s\bigl\langle\boldsymbol{\mathrm{R}}(s)\bigr\rangle-\boldsymbol{\mathrm{r}}(t=0)=\biggl[\langle L\rangle+\frac{g}{\tau}C(s)L_z^2\biggr]\,\bigl\langle\boldsymbol{\mathrm{R}}(s)\bigr\rangle,
\label{laplace}
\end{equation}
where $L_{x/y/z}$ are the 3-dimensional representation of the generators of the algebra of spatial rotations (see Eq. \eqref{so(3)}). The solution of the equation above is:
\begin{equation}
\bigl\langle\boldsymbol{\mathrm{R}}(s)\bigr\rangle=\biggl[s\mathbb{1}_3-\langle L\rangle-\frac{g}{\tau}C(s)L_z^2\biggr]^{-1}\,\boldsymbol{\mathrm{r}}(t=0)\,.
\label{laplace-1}
\end{equation}
\begin{figure}[h]
\centering
\includegraphics[width=\textwidth]{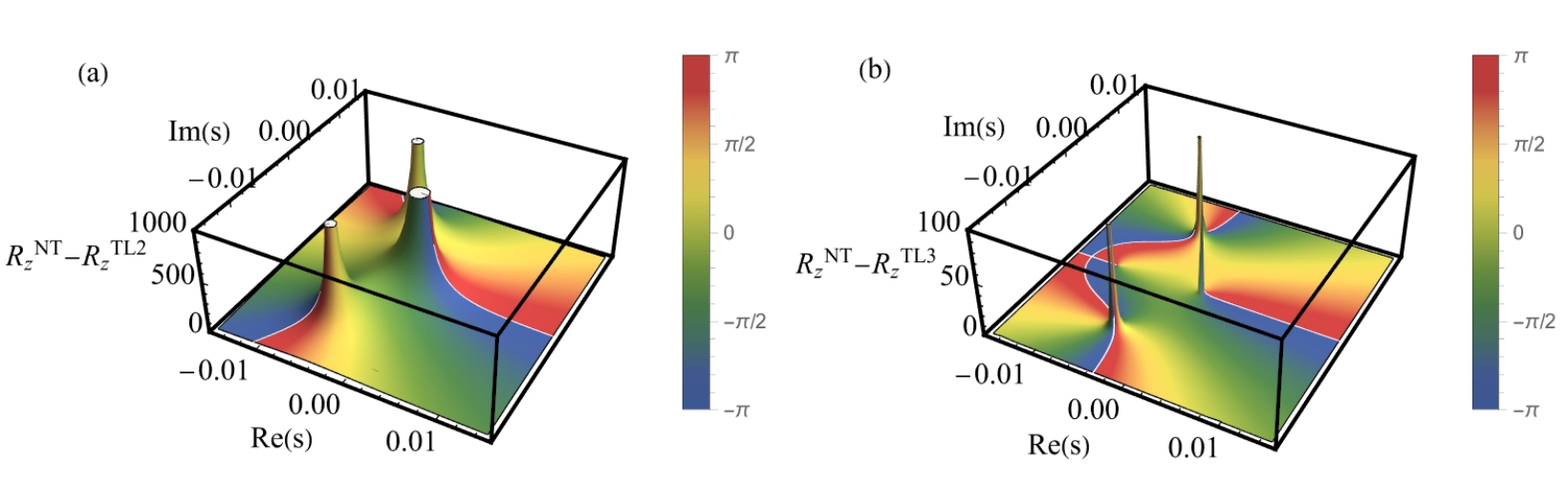}
\caption{
Difference of the Laplace transforms of $\mathrm{R}_z(s)$ obtained from the exact but non-local in time master equation $\mathrm{R}^{\mathrm{NT}}_z(s)$ and the approximate third-order TL one $\mathrm{R}^{\mathrm{TL3}}_z(s)$. In (a) we plot the difference between Eq.~\eqref{laplacezex} and Eq.~\eqref{laplacezII}, while in (b) we plot the difference between Eq.~\eqref{laplacezex} and Eq.~\eqref{laplacezIII} in the same portion of the complex plane. The interesting behaviour of the resulting quantities is concentrated at the poles, as expected. From the comparison of the two plots is evident how the third order is notably closer to the exact solution. Interestingly, also, a smaller difference appears at $s=0$ than in (b) which is due to the memory effect introduced by the third order. This is the improvement due to the inclusion of the third-order, which captures the essentially non-Markovian part of the response.}
\label{LaplaceDiff}
\end{figure}
For OU noise, the Laplace transform can be given analytically. The analytical solution for the $z$ component  is:
\begin{equation}
\mathrm{R}^{\mathrm{NT}}_z(s)=\frac{\Delta^2+\big(s+\frac{g}{1+s\tau}\big)^2}{\Delta^2s+\frac{\big(\tau s^2+s+g\big)\big[\tau s(D^2+ s^2)+s(s+g)+D^2\big]}{1+\tau^2s^2}}\,.
\label{laplacezex}
\end{equation}
The poles of the transform of the correlation function $C(s)$ are visible when the Laplace transform $\mathrm{R}^{\mathrm{NT}}_z(s)$ is plotted in the complex frequency plane.
This is to be compared with the Laplace transform of our time-local equation. We Laplace-transform the second-order time-local (TL) master equation and get to an equation equivalent to Eq.~\eqref{laplace}. Then, before inverting to get the equivalent of Eq.~\eqref{laplace-1}, we discard terms like $\boldsymbol{\mathrm{R}}(1/\tau)$, since $\tau$ is usually small and the Laplace transform has to converge to zero for high frequencies. This can also be checked self-consistently when the solution is given later. After this procedure, we get to:
\begin{equation}
\mathrm{R}^{\mathrm{TL2}}_z(s)=\frac{\Delta^2+\big(s+g\bigr)^2}{\Delta^2s+\big(s+g)\big[D^2+s(s+g)\big]}\,.
\label{laplacezII}
\end{equation}
The expression above is just the memoryless limit of the exact result. In fact, it is just the zeroth-order expansion of Eq.~\eqref{laplacezex} in $\tau$ (i.e., for short memory times).

The third-order correction can be calculated by Laplace-transforming Eq.~\eqref{thirddephasing} and inverting the resulting algebraic equation.
The third-order corrected Laplace transform of the solution of our time-local master equation is:
\begin{equation}
\mathrm{R}^{\mathrm{TL3}}_z(s)=\frac{s\bigl[\Delta^2+(s+g)^2\bigr]}{s\big\{\Delta^2+\big(s+g)\big[D^2+s(s+g)\big]\big\}-gD^2\tau(s+g)}\,.
\label{laplacezIII}
\end{equation}
We note how this expression depends on the memory time of the noise. By plotting the Laplace transform $\mathrm{R}_z(s)$ in the complex frequency plane (plot is not shown for redundancy with Fig.~\ref{LaplaceDiff}), we find that the $\mathrm{R}^{\mathrm{TL3}}_z(s)$ from Eq.~\eqref{laplacezIII} agrees qualitatively very well with the exact $\mathrm{R}_z^{\mathrm{NT}}$ Eq.~\eqref{laplacezex}.
\begin{figure}[h]
 \centering
     \subfloat{\includegraphics[width=0.45\textwidth]{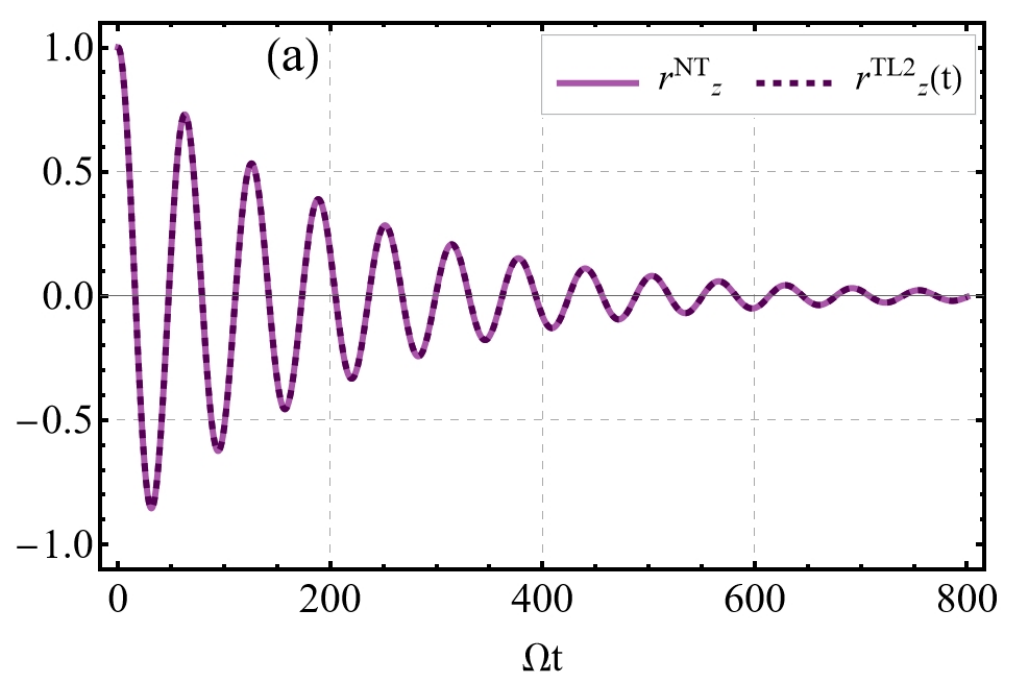}}\,
     \subfloat{\includegraphics[width=0.45\textwidth]{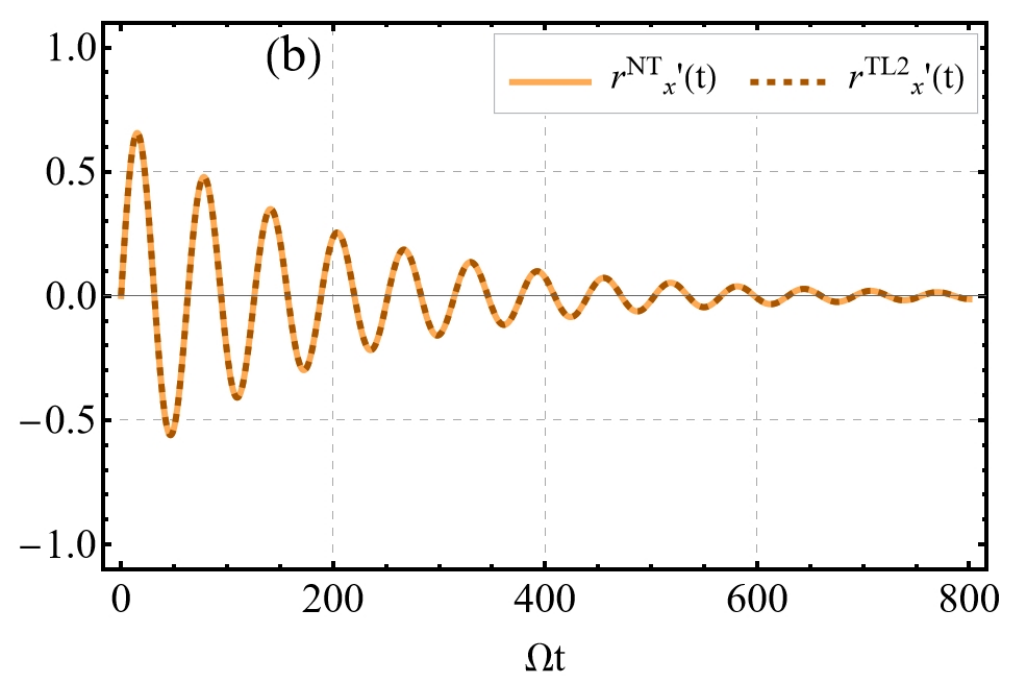}}
\caption{Components of the Bloch vector $\boldsymbol{\mathrm{r}}'(t)=\mathrm {Tr}[\boldsymbol{\sigma}\rho'(t)]$  in the rotating frame for $\varphi=\pi/4\,,\,\tau=0.1\,\Omega^{-1}\,,\,D=10^{-2}\,\Omega\,,\,g=4\times10^{-3}\,\Omega\,$. In (a) we plot the z component, for which $\mathrm{r}_z'=\mathrm{r}_z$ since it is invariant under rotations about $z$. In (b), the x component is plotted. Solid lines represent the exact curves that are solutions of \eqref{eqnovikov} obtained through the inverse Laplace transform. Dashed lines represent the approximate solutions from the time-local master equation.}
\label{BlochTraj2nd}
\end{figure}

As is evident from the analytic expressions above, the third order brings into the linear response the contribution that is essentially non-Markovian, i.e., the part of the response of the system that is aided by the memory of the bath. We recover the Markovian case Eq.~\eqref{laplacezII} by setting $\tau=0$. To show the good improvement we obtain with the third-order term, we plot the difference between \eqref{laplacezex} and \eqref{laplacezII} in Fig.~\ref{LaplaceDiff}(a), while in Fig.~\ref{LaplaceDiff}(b) we plot the difference between \eqref{laplacezex} and \eqref{laplacezIII} in the complex frequency plane, computed assuming $\boldsymbol{\mathrm{r}}(t=0)=[0\,\,0\,\,1]^T$. The accuracy of the third-order-corrected master equation is indeed striking, as can be seen from the different scales on the z-axis of the graph, and the smallness of the error made in approximating the $|s|=0$ part of the function.
\begin{figure}[h]
 \centering
     \subfloat{\includegraphics[width=0.45\textwidth]{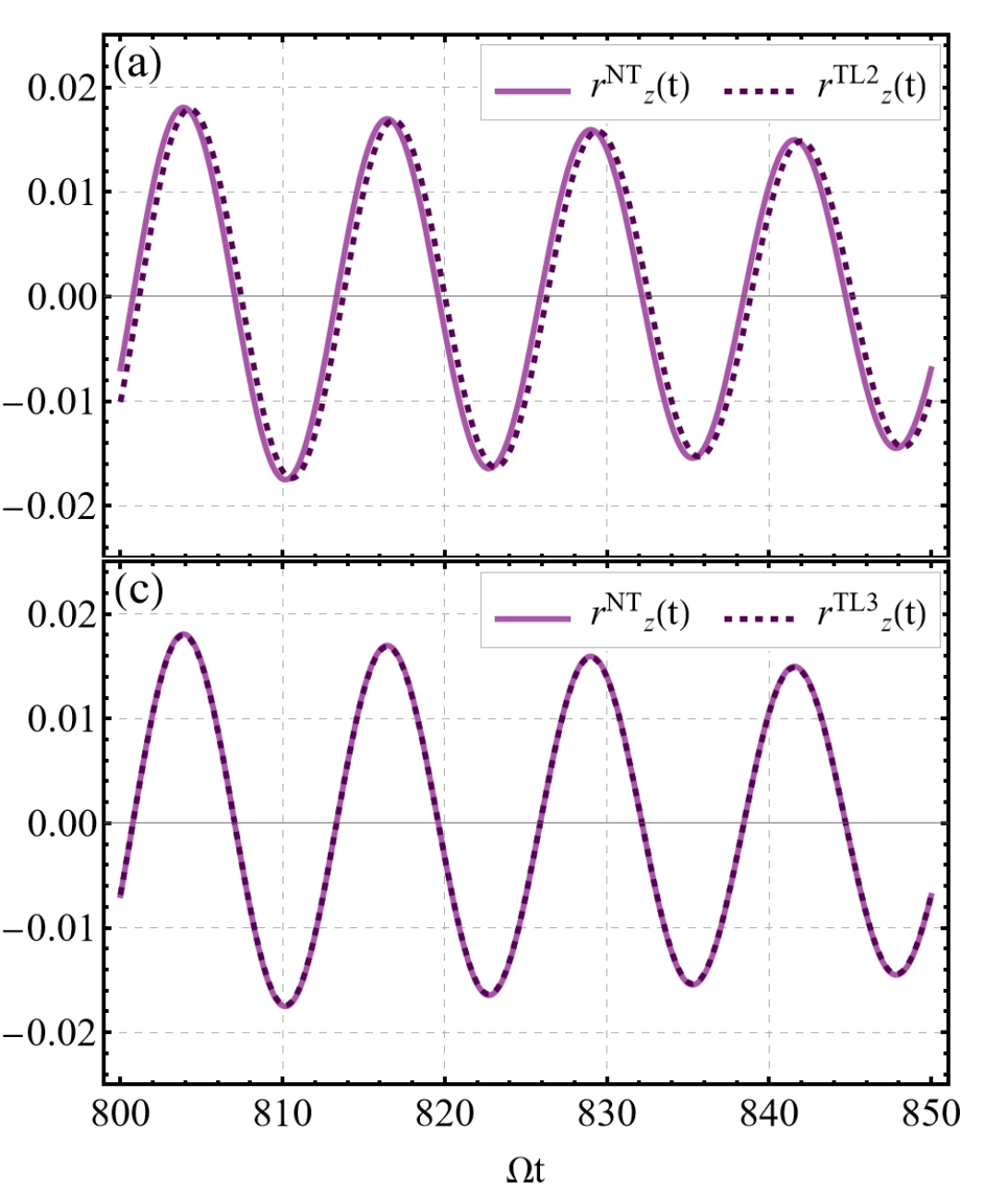}}\,
     \subfloat{\includegraphics[width=0.45\textwidth]{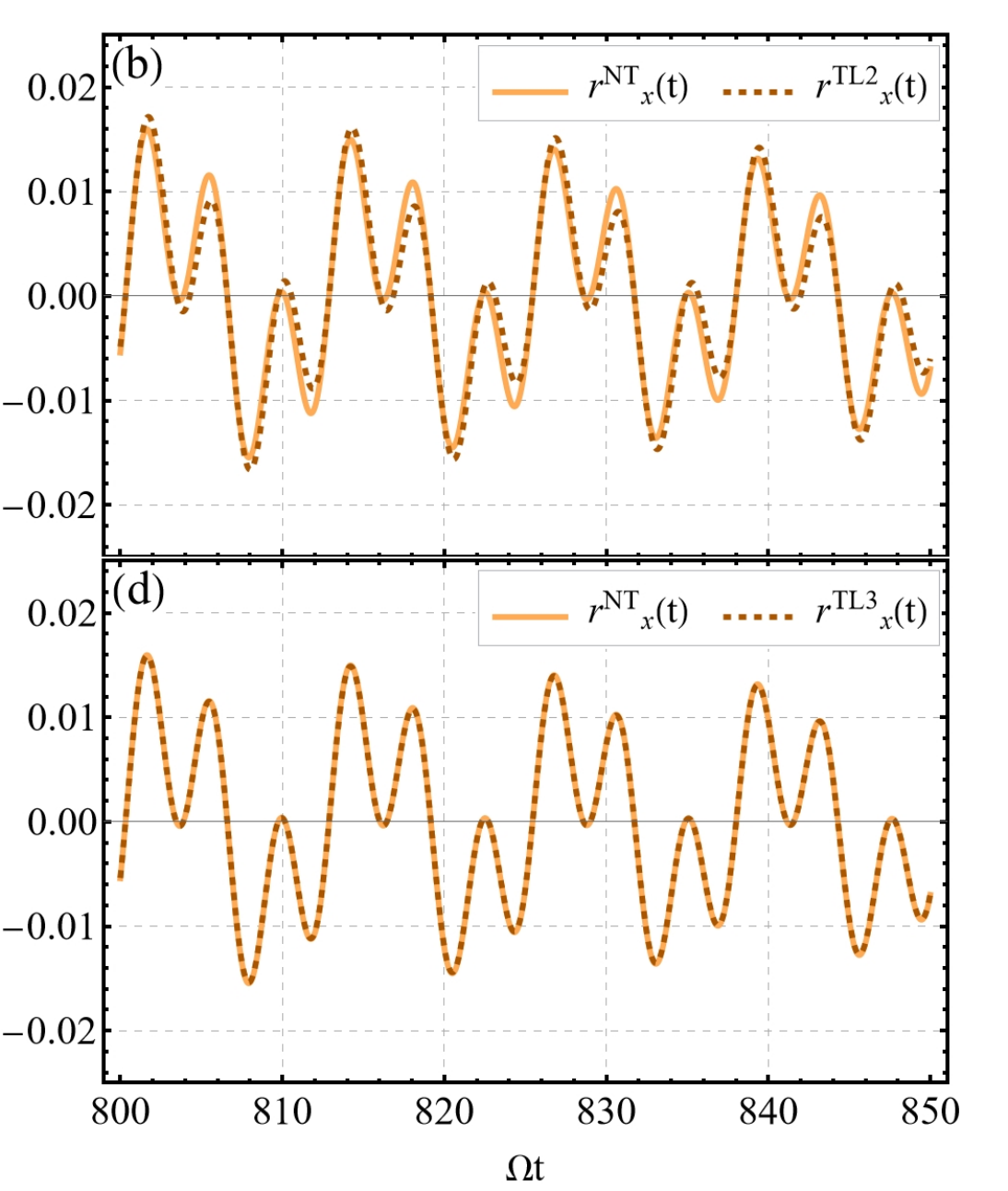}}
\caption{Long-time behaviour of our TL approximation for $\varphi=\pi/4\,,\,\tau=0.1\,\Omega^{-1}\,,\,D=5\times10^{-2}\,\Omega\,,\,g=4\times10^{-3}\,\Omega\,$.  Components of the Bloch vector $\boldsymbol{\mathrm{r}}(t)={\rm Tr}[\boldsymbol{\sigma}\rho(t)]$ are plotted. Solid lines represent the exact curves that are solutions of Eq.~\eqref{eqnovikov} obtained through the inverse Laplace transform. Dashed lines represent the approximate solutions from the second-order time-local master equation in (a) and (b) and from the third-order time-local master equation in (c) and (d). In (a) and (c), the $z$ component of the Bloch vector is plotted, while in (b) and (d) the $x$ component is plotted. It is evident that, going from the upper panels to the bottom ones, the accuracy is improved thanks to the inclusion of the third-order term in the master equation.}
\label{BlochTraj}
\end{figure}

By performing the inverse Laplace transform on the solution of Eq.~\eqref{laplace}, we can produce exact solutions in the time domain, see Fig.~\ref{BlochTraj2nd}. For ease of visualization, we also consider the equation in the rotating frame and plot an alternative vectorized form $\boldsymbol{\mathrm{r}}'(t)=\mathrm {Tr}[\boldsymbol{\sigma}\rho'(t)]$ of the Bloch vector, which compensates for the additional, nonessential frequency components. We call $\boldsymbol{\mathrm{r}}'$ the Bloch vector, nonetheless, even if its components (apart from $\mathrm{r}_z'=\mathrm{r}_z$) are not expectation values. 
In Fig.~\ref{BlochTraj2nd} we show the Laplace inverse of $\boldsymbol{\mathrm{R}}'(s)$ which is the Bloch vector $\boldsymbol{\mathrm{r}}'(t)$ \cite{Feynman57}.
In the following, we set $\varphi=\pi/4$. As is evident from Fig.~\ref{BlochTraj2nd}, the corresponding solid and dashed curves match perfectly, meaning that our approximation works well in the weak-coupling and weak-driving limit.

To go to stronger driving regimes, we include the third-order cumulant in our description. Using the expression of Eq.~\eqref{thirddephasing}, we obtain the corresponding term written Bloch vector (BV) basis, in the rotating frame:
\begin{equation}
\mathbb{K'}^{\rm{III}}(t)=\frac{gD}{\tau}\int_0^tdt'\int_0^{t-t'}dt''\,c(t'+t'')
\begin{bmatrix}
0 && 0 && \sin[\varphi]\\
0 && 0 && \cos[\varphi]\\
0 && 0 && 0\\
\end{bmatrix}\,.
\end{equation}

We show the result of the application of the third-order TL master equation to the dynamics of Bloch vector components
in Fig.~\ref{BlochTraj} in the long-time limit, where failures of the second-order description are expected. As can be seen in Fig.~\ref{BlochTraj}, the third order corrects the long-time behavior of the standard second-order approximation.


\section{Qubit in a quantum environment}
\label{sec:qubitrelax}

We next demonstrate the accuracy of Eq.~\eqref{meqquantum} in the particular example of a driven qubit in the presence of a Bosonic bath. We assume that the bath consists of a collection of Bosonic modes, $\mathrm{H_B}=\sum_k\omega_ka_k^\dagger a_k$, that interacts with a driven qubit, described by $\mathrm{H}_S(t)=\Omega\sigma_+\sigma_-+\mathrm{V}(t)$, through energy exchange in the RWA, that is, $\mathrm{H}_{BS}=\sum_kg_k(\sigma_+a_k+\sigma_-a_k^\dagger)\equiv\sigma_+B+\sigma_-B^\dagger$. At zero temperature, the bath is considered to be initially in its vacuum state $\rho_B=|0\rangle\langle0|_B$ so that its only nonzero correlation function is $c(t)=\langle B(t)B^\dagger(0)\rangle_B=\sum_k|g_k|^2e^{-i\omega_kt}$. In the continuum limit, the sum over the coupling constants is replaced with an integral weighted by the density of states of the reservoir modes, i.e., the spectral density: $c(t)=\int_{-\infty}^\infty\mathrm d\omega\ D(\omega)e^{-i\omega t}$. Given that this integral can be evaluated using complex contour integration techniques, we can introduce pseudo-modes (PMs) into the system to exactly describe the arising memory effects \cite{Garr1,Garr2,Garr3,Damanet2024}. Moreover, PMs constitute a reliable approximation scheme even for the finite temperature case \cite{Eisfeld2014}.

\begin{figure}[t]
\begin{center}\includegraphics[width=0.9\textwidth]{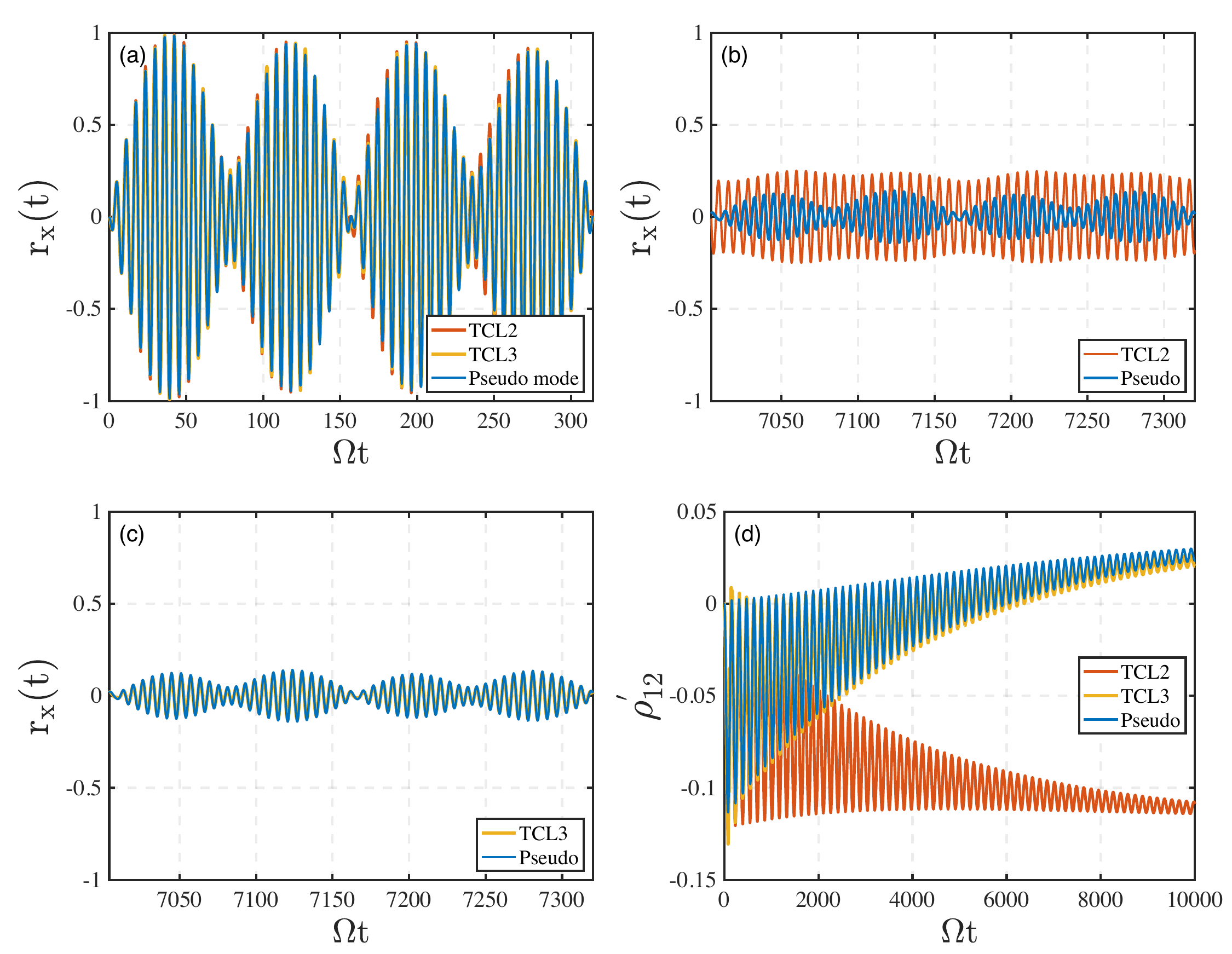}
\end{center}
 \caption{Driven qubit in a quantum environment with a single PM (N=1) corresponding to a Lorentzian power spectrum. (a) The short-time dynamics of the Bloch vector component $\mathrm{r_x}$ showing a beating pattern. The second (red line) and third order (yellow line) approximations agree well with the exact pseudo mode (blue line) description. (b) The long-time dynamics of the same Bloch vector component reveal a decaying beating pattern in the exact solution, which is not captured by the TCL2 approximation. (c) The long-time dynamics is approximated very well with the inclusion of the third order correction. (d) The dynamics of the real part of the off-diagonal element of the density matrix in the interaction picture reveal the feature that is not captured by the second-order approximation. The parameters used for these plots are $\eta/\Omega=0.035,\xi/\Omega=0.75,\Gamma/\Omega=0.02,D/\Omega=0.04$. }
\label{pse}
\end{figure}

Assuming that the spectral density of the environment $D(\omega)$ tends to zero at least as fast as $\mathcal O(1/|\omega|^2)$ for $|\omega|\to\infty$ (in some relevant cases requiring a high-frequency cutoff) and that it lacks non-analytical features such as a branch cut, the two-time correlation function $c(t)$ can be fully described by the poles and residues of the spectral density \cite{Garr3}. By the residue theorem, we write the Bosonic bath's correlation function as
\begin{equation}
    c(t)=-i\Lambda_0^2\sum_{l=1}^Nr_le^{-iz_lt},\label{eq:corrfunc}
\end{equation}
where $\Lambda_0$ is an overall coupling strength and $z_l=\xi_l-i\Gamma_l/2$ are the poles of the spectral density located only in the lower half of the complex plane, i.e, $\Gamma_l>0$. The poles have respective residues $r_l$, such that $-ir_l>0$. To check this, one can evaluate $c(0)=\sum_k|g_k|^2$ that is a real positive number. Therefore, from Eq. \eqref{eq:corrfunc}, $-i\Lambda_0^2\sum_{l=1}^Nr_l$ must be also real and positive. Then, $-ir_l$ is a real number and we restrict ourselves to positive cases, so that the effective couplings $\eta_l$, introduced later in Eq.~\eqref{eq:pseudohamilton}, of the PMs are real. This treatment can be extended straightforwardly to complex PM coupling constants when needed \cite{Garr3}.
The PMs consist of $N$ Bosonic modes associated with the poles. They are described by the annihilation(creation) operators $b_l^{(\dagger)}$, whose energy $\xi_l$ and decay rates $\Gamma_l$ are determined by the real and imaginary parts of the poles of the correlation function $c(t)$, respectively. The exact master equation governing the dynamics of the combined state $\rho(t)$ of the driven qubit along with the PMs is \cite{Garr3}
\begin{gather}
    \dot\rho(t)=-i[\mathrm{H}(t),\rho(t)]+\sum_l\Gamma_l\Bigl[b_l\rho(t)b_l^\dagger-\frac{1}{2}\{b_l^\dagger b_l,\rho(t)\}\Bigr],\nonumber\\
    \mathrm{H}(t)=\mathrm{H}_S(t)+\sum_l\xi_lb_l^\dagger b_l+\sum_l\eta_l\bigl(\sigma_+b_l+\sigma_-b_l^\dagger\bigr),\label{eq:pseudohamilton}
\end{gather}
where the coupling strengths between the qubit and the PMs are $\eta_l=\Lambda_0\sqrt{-ir_l}$. The exact reduced dynamics of the qubit state is obtained after tracing over the PM degrees of freedom $\rho_S(t)=\mathrm{Tr}_{\mathrm{PM}}\rho(t)$. 

At this point, we mention that in the continuum limit a Bosonic environment following the fluctuation-dissipation theorem would have $D(\omega<0)=0$ at zero temperature. As a result, no finite number of PMs can exactly reproduce the two-time correlation function of a continuous Bosonic environment~\cite{Plenio2018,Lambert2019,Stockburger2022}. However, we emphasize that the goal of this example is to show the accuracy of our approximation using an analytically tractable model, analogously to the approach adopted in Ref.~\cite{Hartmann2020}. This simplified model assumes the bath correlation function of Eq.~\eqref{eq:corrfunc} together with an initial Gaussian state of the environment, thereby fulfilling the assumptions of the theorem in Ref.~\cite{Plenio2018}, according to which the reduced dynamics obtained from Eq.~\eqref{eq:pseudohamilton} is equivalent to the exact non-Markovian open quantum system dynamics of the reduced system interacting with the continuous Bosonic bath. The bath correlation function may then be expressed as a finite sum of exponentials, e.g., through a fitting procedure, with a finite approximation error that can be systematically reduced, as discussed in Ref.~\cite{Lambert2019}.

The exact description provides an ideal benchmark for the approximate evolution in Eq.~\eqref{meqquantum} along with the third-order correction from Eq.~\eqref{eqbathdrivethird}. Evaluating these, we find the approximate master equation in the Schrödinger picture,
\begin{gather}
    \dot\rho_S(t)=-i\bigl[[\Omega+\Gamma_I(t)]\sigma_+\sigma_-+\mathrm V(t),\rho_S(t)\bigr]+2\Gamma_R(t)\Bigl[\sigma_-\rho_S(t)\sigma_+-\frac{1}{2}\{\sigma_+\sigma_-,\rho_S(t)\}\Bigr]\nonumber\\-ig(t)\bigl[\sigma_z\rho_S(t)\sigma_++\sigma_+\rho_S(t)\bigr]+ig^*(t)\bigl[\sigma_-\rho_S(t)\sigma_z+\rho_S(t)\sigma_-\bigr],\label{psemeq}
\end{gather}
where $\Gamma_{R/I}(t)$ is the real/imaginary part of $\Gamma(t)=\int_0^t\mathrm{d}t_1\ e^{i\Omega t_1}c(t_1)$. In the equation above, incidentally, the first line of the RHS comes from the second cumulant and the second line comes from the third cumulant. For the third order term, we assumed the driving term has the form $\mathrm{V}(t)=f(t)\sigma_++f^*(t)\sigma_-$, hence the function appearing as the third order rate is
\begin{equation}
    g(t)=\int_0^tdt_1\int_0^{t_1}dt_2\ f(t_1)c(t-t_2)e^{i\Omega(t_1-t_2)}.
\end{equation}
For consistency with the discussion in Section~\ref{sec:noisy}, by expressing the density matrix as $\rho(t)=\bigl[\mathbb 1+\mathbf r(t)\cdot\boldsymbol{\sigma}\bigr]/2$, we rewrite Eq.~\eqref{psemeq} for the BV:
\begin{equation}
    \frac{d}{dt}\begin{bmatrix}
        \mathrm{r_x}\\
        \mathrm{r_y}\\
        \mathrm{r_z}
    \end{bmatrix}=\begin{bmatrix}
        -\Gamma_R(t) & -\Omega-\Gamma_I(t) & -2f_I(t)\\
        \Omega+\Gamma_I(t) & -\Gamma_R(t) & -2f_R(t) \\
        2[f_I(t)+g_I(t)] & 2[f_R(t)+g_R(t)] & -2\Gamma_R(t)
    \end{bmatrix}\cdot\begin{bmatrix}
        \mathrm{r_x}\\
        \mathrm{r_y}\\
        \mathrm{r_z}
    \end{bmatrix}+2\begin{bmatrix}
        g_I(t)\\
        g_R(t)\\
        -\Gamma_R(t)
    \end{bmatrix},
\label{BVeq}
\end{equation}
where $f_{R/I}$ and $g_{R/I}$ denote the real/imaginary part of the drive function and the third-order rate. As is evident from this formulation, the third-order term $g(t)$ appears in the inhomogeneous term, thereby modifying the long-time behavior of the dynamics. Interestingly, $g(t)$ appears asymmetrically in the Bloch matrix, influencing the drive term only for the equation of $\mathrm{r_z}$.

In Fig.~\ref{pse} we show the dynamics of the Bloch vector component along the $x$ axis for resonant circularly polarized driving, i.e., $f(t)=\frac{D}{2}\,e^{-i\Omega t}$. We considered a single PM, corresponding to a continuum with a Lorentzian power spectrum. This can be seen as a Jaynes-Cummings model with a leaky cavity. We find that the third-order TCL equation accurately captures the dynamics described by the exact PM master equation. In contrast, the second-order TCL equation fails to reproduce the correct long-time behavior.

\section{Conclusions}
We have presented a new and more formal derivation of the TL master equations, using an approach to perturbation theory where both the time-dependent drive and the dissipative interactions are treated as perturbations on the same footing. The master equation is not derived, as is usually done, by going into the interaction picture following the eigenstates of the full coherent dynamics. Instead, we treat both coherent and incoherent terms perturbatively, going into the interaction picture with respect to the unperturbed eigenbasis of the bare Hamiltonian. Of course, this is allowed only when the strengths of the perturbations are smaller than the level spacing of the unperturbed Hamiltonian. This allows us to derive a relatively simple TL master equation. The second-order truncation of this equation, corresponding to the Born approximation, indeed gives the Bloch-Redfield theory. This justifies the usual treatment of such problems, where the dissipator is typically computed by neglecting the drive term and subsequently inserting it into the von Neumann part of the equation.

Going beyond this approximation, we have derived the first correction in the drive strength, in what we call the third-order master equation, which is still second-order in the bath-system coupling and first order in the drive strength, and therefore must be included in the usual Bloch-Redfield theory of relaxation. We showed how this third-order term notably improves the agreement with exact solutions of simple scenarios. The improvement is particularly good for the dynamics of the coherences of the reduced state.
Quite interestingly, the third order in the quantum noise case predicts a striking effect. In fact, the last term in Eq.~\eqref{BVeq} suggests that it may be possible to dissipatively engineer a target steady state (even one with non-zero coherences) by coherent pulses. This unleashes unprecedented capabilities in coherent control. An immediate application can be the design of a protocol for a \emph{coherent} Overhauser effect, meaning the partial (limited-fidelity) transfer of the quantum state from a controlled spin to another, uncontrolled, one, only through the interaction with a shared environment. The design of this protocol should be quite straightforward thanks to the simple form of equation \eqref{BVeq}. Therefore, mimicking already existing protocols for the \emph{incoherent} Overhauser effect \cite{SlichterBK,DiBari92,Ernst95}, i.e., the one involving only the transfer of the populations of the spins, might be sufficient. With similar protocols, one could engineer coherent pulses for producing a steady state with some quantum correlations in a collection of spins by controlling individual spins that are only virtually coupled \cite{DiBari92,Ernst95}.

The simple examples we discussed were designed to benchmark our treatment with analytically solvable cases. Nonetheless, the great benefit of these methods will be showcased in more complex problems, which are the subject of future works. The striking improvement of old approximation techniques makes us confident that our approach will be useful for the design of improved quantum control schemes, which are relevant for the quantum-optics and quantum-information-processing communities \cite{Petta2024,Scarlino2025,Lidar2025,Andersen2025,Gasparinetti2025}.

We have shown that the procedure of elimination of the drive from the second-order dissipator in this alternative perturbation theory framework can be performed irrespective of the classical or quantum nature of the fluctuations. Thus, we say that this fact pertains to the dynamics encoded in the TL master equation structure.

Finally, it is worth emphasizing that this approach extends the linear response of the system in the non-Markovian case. The results then derived, including the third-order term of the master equation, can be of use if one wants to guess the linear response of a driven many-body system. Moreover, the third order presents such a simple dependence on the drive function that applications to solid-state physics, and ultra-fast atomic and quantum optics might be envisioned. It might be important to study from this perspective the implications of the phenomenology of some quantum processes that we are now able to investigate in the time domain. In particular, this theory could be important for the phenomenology of quantum stochastic resonance \cite{Grifoni2000,Flindt2009,Haug2019,Haug2025} and resonance fluorescence \cite{Mollow69,Mork2012,Hughes2018,Mueller2024,Stobinska2024}. The application of the present theory to the driven-dissipative dynamics of Bosonic modes is also worthy of further investigation and will be the subject of future work.

\ack
L.B. and G.B. gratefully acknowledge funding from the Deutsche
Forschungsgemeinschaft (DFG, German Research Foundation) -- Project No. 425217212 --
SFB 1432.
B.G. and G.B. acknowledge the support from the State of Baden-W\"urttemberg within the Competence Center Quantum Computing, project KQCBW24.

\appendix

\section{Fourth order}
\label{appendixiv}

The fourth-order term of the cumulant expansion is given by \cite{Fox76}:
\begin{align}
\langle\langle\mathcal{L}'(t)\mathcal{L}'(t_1)\mathcal{L}'(t_2)\mathcal{L}'(t_3)\rangle\rangle&=\langle\mathcal{L}'(t)\mathcal{L}'(t_1)\mathcal{L}'(t_2)\mathcal{L}'(t_3)\rangle-\langle\mathcal{L}'(t)\rangle\langle\mathcal{L}'(t_1)\mathcal{L}'(t_2)\mathcal{L}'(t_3)\rangle\nonumber\\
&-\langle\mathcal{L}'(t)\mathcal{L}'(t_1)\mathcal{L}'(t_2)\rangle\langle\mathcal{L}'(t_3)\rangle
-\langle\mathcal{L}'(t)\mathcal{L}'(t_1)\mathcal{L}'(t_3)\rangle\langle\mathcal{L}'(t_2)\rangle\nonumber\\
&-\langle\mathcal{L}'(t)\mathcal{L}'(t_2)\mathcal{L}'(t_3)\rangle\langle\mathcal{L}'(t_1)\rangle
-\langle\mathcal{L}'(t)\mathcal{L}'(t_1)\rangle\langle\mathcal{L}'(t_2)\mathcal{L}'(t_3)\rangle\nonumber\\
&-\langle\mathcal{L}'(t)\mathcal{L}'(t_2)\rangle\langle\mathcal{L}'(t_1)\mathcal{L}'(t_3)\rangle-\langle\mathcal{L}'(t)\mathcal{L}'(t_3)\rangle\langle\mathcal{L}'(t_1)\mathcal{L}'(t_2)\rangle\nonumber\\
&+\langle\mathcal{L}'(t)\rangle\langle\mathcal{L}'(t_1)\rangle\langle\mathcal{L}'(t_2)\mathcal{L}'(t_3)\rangle+\langle\mathcal{L}'(t)\rangle\langle\mathcal{L}'(t_2)\rangle\langle\mathcal{L}'(t_1)\mathcal{L}'(t_3)\rangle\nonumber\\
&+\langle\mathcal{L}'(t)\rangle\langle\mathcal{L}'(t_3)\rangle\langle\mathcal{L}'(t_1)\mathcal{L}'(t_2)\rangle+\langle\mathcal{L}'(t)\rangle\langle\mathcal{L}'(t_1)\mathcal{L}'(t_2)\rangle\langle\mathcal{L}'(t_3)\rangle\nonumber\\
&+\langle\mathcal{L}'(t)\rangle\langle\mathcal{L}'(t_1)\mathcal{L}'(t_3)\rangle\langle\mathcal{L}'(t_2)\rangle+\langle\mathcal{L}'(t)\rangle\langle\mathcal{L}'(t_2)\mathcal{L}'(t_3)\rangle\langle\mathcal{L}'(t_1)\rangle\nonumber\\
&+\langle\mathcal{L}'(t)\mathcal{L}'(t_1)\rangle\langle\mathcal{L}'(t_2)\rangle\langle\mathcal{L}'(t_3)\rangle+\langle\mathcal{L}'(t)\mathcal{L}'(t_2)\rangle\langle\mathcal{L}'(t_1)\rangle\langle\mathcal{L}'(t_3)\rangle\nonumber\\
&+\langle\mathcal{L}'(t)\mathcal{L}'(t_3)\rangle\langle\mathcal{L}'(t_1)\rangle\langle\mathcal{L}'(t_2)\rangle+\langle\mathcal{L}'(t)\mathcal{L}'(t_1)\rangle\langle\mathcal{L}'(t_3)\rangle\langle\mathcal{L}'(t_2)\rangle\nonumber\\
&+\langle\mathcal{L}'(t)\mathcal{L}'(t_2)\rangle\langle\mathcal{L}'(t_3)\rangle\langle\mathcal{L}'(t_1)\rangle+\langle\mathcal{L}'(t)\mathcal{L}'(t_3)\rangle\langle\mathcal{L}'(t_2)\rangle\langle\mathcal{L}'(t_1)\rangle\nonumber\\
&-\langle\mathcal{L}'(t)\rangle\langle\mathcal{L}'(t_1)\rangle\langle\mathcal{L}'(t_2)\rangle\langle\mathcal{L}'(t_3)\rangle\nonumber-\langle\mathcal{L}'(t)\rangle\langle\mathcal{L}'(t_1)\rangle\langle\mathcal{L}'(t_3)\rangle\langle\mathcal{L}'(t_2)\rangle\\
&-\langle\mathcal{L}'(t)\rangle\langle\mathcal{L}'(t_2)\rangle\langle\mathcal{L}'(t_1)\rangle\langle\mathcal{L}'(t_3)\rangle\nonumber-\langle\mathcal{L}'(t)\rangle\langle\mathcal{L}'(t_2)\rangle\langle\mathcal{L}'(t_3)\rangle\langle\mathcal{L}'(t_2)\rangle\\
&-\langle\mathcal{L}'(t)\rangle\langle\mathcal{L}'(t_3)\rangle\langle\mathcal{L}'(t_1)\rangle\langle\mathcal{L}'(t_2)\rangle\nonumber-\langle\mathcal{L}'(t)\rangle\langle\mathcal{L}'(t_3)\rangle\langle\mathcal{L}'(t_2)\rangle\langle\mathcal{L}'(t_1)\rangle ,
\end{align}
where $\mathcal{L}'(t)=\mathcal{L}_s'(t)+\mathcal{L}_d'(t)$.

One can show with some algebra that there is no fourth-order term in the drive strength. When assuming Gaussian noise, it follows that there can be no contributions which are first or third order in the drive $\mathcal{L}_d'(t)$, since those will leave a $\mathcal{L}_s'(t_i)$ unpaired, leading to a null average. There are then terms which are second order in the drive strength. Clearly, there is also the fourth-order dissipator. Then we see that the lengthy expression for the cumulant above can be rewritten in the following, relatively simple, form:
\begin{align}
\langle\langle\mathcal{L}'(t)\mathcal{L}'(t_1)\mathcal{L}'(t_2)\mathcal{L}'(t_3)\rangle\rangle&=\langle\mathcal{L}_s'(t)\mathcal{L}_s'(t_1)\mathcal{L}_s'(t_2)\mathcal{L}_s'(t_3)\rangle-\langle\mathcal{L}_s'(t)\mathcal{L}_s'(t_1)\rangle\langle\mathcal{L}_s'(t_2)\mathcal{L}_s'(t_3)\rangle\nonumber\\
&-\langle\mathcal{L}_s'(t)\mathcal{L}_s'(t_2)\rangle\langle\mathcal{L}_s'(t_1)\mathcal{L}_s'(t_3)\rangle-\langle\mathcal{L}_s'(t)\mathcal{L}_s'(t_3)\rangle\langle\mathcal{L}_s'(t_1)\mathcal{L}_s'(t_2)\rangle\nonumber\\
&+\langle\mathcal{L}_s'(t)\mathcal{L}_d'(t_1)\mathcal{L}_d'(t_2)\mathcal{L}_s'(t_3)\rangle\,.
\label{IVcumulant}
\end{align}
The first four terms represent the fourth-order dissipator, which is neglected under the Born approximation. This requires $\Omega,\delta\gg\eta^2\tau$, where $\Omega$ is the level spacing of the system, $\delta$ characterizes the drive strength, $\eta$ is the noise strength or system-bath coupling and $\tau$ is the bath correlation time, which gives a rough estimate of the effective integration interval of the dissipator, i.e., the time over which the bath correlation function dies out. The fifth term on the RHS of Eq. \eqref{IVcumulant} corresponds to the correction to the third-order dissipator, accurate up to second order in the drive. Assuming a harmonic drive, the third-order dissipator is the leading non-vanishing contribution, which can be roughly estimated as $\eta^2\tau\delta/\omega$, where $\omega$ is the drive frequency. To ensure its dominance over higher-order corrections, we require $\eta^2\tau\delta/\omega\gg\eta^2\tau(\delta/\omega)^2$, which effectively constitutes a form of secular approximation, albeit one more sophisticated than the usual. The rates that we are comparing in this way are the number of photons introduced by the drive per unit of time $\propto\delta$ and the frequency of the drive $\omega$, whose inverse gives the effective extension of the integrals of the coherent parts. Remarkably, the effectiveness of the third-order dissipator, evident from the figures in this paper, comes from the fact that it contains the qualitative features of processes assisted by the memory of the bath. In fact, it captures the first-order processes in the drive photons, assisted by the memory of the environment. The fifth term in Eq.~\eqref{IVcumulant}, then, would capture the two-drive-photon processes. Higher-order processes in the drive will presumably appear in higher cumulants.

\section{Explicit calculations with the projection operator}
\label{sec:AppendixBath}
Here, we expand $\mathbb{K}_q(t)$ from the master equation in Eq.~\eqref{eqshibata} in a series up to second order in $\mathcal{L}'$,
\begin{equation}
\mathbb{K}_q(t)=-i\mathcal{P}\mathcal{L}'(t)-\int_0^t \bigl[\mathcal{P}\mathcal{L}'(t)\mathcal{L}'(t-t')-\mathcal{P}\mathcal{L}'(t)\mathcal{P}\mathcal{L}'(t-t')\bigr]dt'.
\label{firstappendix}
\end{equation}

Now we can apply the same result as with classical noise. Namely, we separate $\mathcal{L}'(t)=\mathcal{L}_D'+\mathcal{L}_{BS}'$ where $\mathcal{L}_D$ depends on the driving and therefore only on the system's coordinates and $\mathcal{L}_{BS}$ depends also on the bath DoFs through the bath-system interaction Hamiltonian.
In a more general way, we can prove a similar statement starting from Eq.~\eqref{firstappendix} and making some fairly plausible assumptions on the projector. From the additivity of the starting Hamiltonian and the fact that the Hilbert space is a direct product, it follows that $\mathcal{L}=\mathcal{L}_0+\mathcal{L}_D+\mathcal{L}_B+\mathcal{L}_{BS}=\mathcal{L}_S+\mathcal{L}_B+\mathcal{L}_{BS}$. Assuming then $\mathcal{P}\mathcal{L}_B=\mathcal{L}_B\mathcal{P}=0$, $\mathcal{P}\mathcal{L}_S=\mathcal{L}_S\mathcal{P}$ 
and $\mathcal{P}\mathcal{L}_{BS}\mathcal{P}=0$ \cite{VanKampenBK}, i.e., that the bath is Gaussian in the sense of classical stochastic processes and has zero average, we have that Eq.~\eqref{firstappendix} gives (using the linearity of the projector) \cite{Chaturvedi79,Petruccione99,Breuer2007,Lidar2007}:
\begin{align}
\mathcal{P}\dot{\rho}'&=\mathbb{K}_q(t)\mathcal{P}\rho'(t)=\label{secondappendix}\\
&\Bigl\{-i\mathcal{P}\mathcal{L}'(t)-\int \bigl[\mathcal{P}\mathcal{L}'(t)\mathcal{L}'(t')-\mathcal{P}\mathcal{L}'(t)\mathcal{P}\mathcal{L}'(t')\bigr]dt'\,\Bigl\}\mathcal{P}\rho'(t)\nonumber\\
=&\Bigl\{-i\mathcal{P}e^{-i(\mathcal{L}_0+\mathcal{L}_B)t}\mathcal{L}(t)e^{i(\mathcal{L}_0+\mathcal{L}_B)t}\nonumber\\
&-\int \bigl[\mathcal{P}e^{-i(\mathcal{L}_0+\mathcal{L}_B)t}\mathcal{L}(t)e^{i(\mathcal{L}_0+\mathcal{L}_B)t'}\mathcal{L}(t-t')e^{i(\mathcal{L}_0+\mathcal{L}_B)(t-t')}\nonumber\\
&-\mathcal{P}e^{-i(\mathcal{L}_0+\mathcal{L}_B)t}\mathcal{L}(t)e^{i(\mathcal{L}_0+\mathcal{L}_B)t}\mathcal{P}e^{-i(\mathcal{L}_0+\mathcal{L}_B)(t-t')}\mathcal{L}(t-t')e^{i(\mathcal{L}_0+\mathcal{L}_B)(t-t')}\bigr]dt'\,\Bigl\}\mathcal{P}\rho'(t)
\nonumber\\
=&\Bigl\{-ie^{-i\mathcal{L}_0t}\mathcal{L}_D(t)e^{i\mathcal{L}_0t}\nonumber\\
&-\int e^{-i\mathcal{L}_0t}\mathcal{P}\mathcal{L}_{BS}(t)e^{i(\mathcal{L}_0+\mathcal{L}_B)t'}\mathcal{L}_{BS}(t-t')e^{i(\mathcal{L}_0+\mathcal{L}_B)(t-t')}dt'\,\Bigl\}\mathcal{P}\rho'(t)\,,\nonumber
\end{align}
where the Liouvillean in the intermediate step is the full Liouvillean minus the bare part and we omitted the integration bounds $\{0\,,\,t\}$ in all integrals to ease the notation.
In this way, we obtain the following master equation in the Schr\"odinger picture,
\begin{equation}
\mathcal{P}\dot{\rho}=-i\mathcal{L}_S(t)\mathcal{P}\rho-\int_0^t\mathcal{P}\mathcal{L}_{BS}(t)e^{i(\mathcal{L}_0+\mathcal{L}_B)t'}\mathcal{L}_{BS}(t-t')e^{-i(\mathcal{L}_0+\mathcal{L}_B)t'}dt'\mathcal{P}\rho\,.
\label{finalappendix}
\end{equation}
As is evident, the drive has dropped from the dissipator.

\section*{References}
\bibliographystyle{iopart-num} 
\bibliography{references} 

\providecommand{\newblock}{}
\begin{thebibliography}{10}
\expandafter\ifx\csname url\endcsname\relax
  \def\url#1{{\tt #1}}\fi
\expandafter\ifx\csname urlprefix\endcsname\relax\def\urlprefix{URL }\fi
\providecommand{\eprint}[2][]{\url{#2}}

\bibitem{Schlosshauer2019}
Schlosshauer M 2019 {\em Phys. Rep.\/} {\bf 831} 1 ISSN 0370-1573
  \urlprefix\url{https://www.sciencedirect.com/science/article/pii/S0370157319303084}

\bibitem{BurkardRMP2023}
Burkard G, Ladd T~D, Pan A, Nichol J~M and Petta J~R 2023 {\em Rev. Mod.
  Phys.\/} {\bf 95}(2) 025003
  \urlprefix\url{https://link.aps.org/doi/10.1103/RevModPhys.95.025003}

\bibitem{Burkard2008}
Chirolli L and Burkard G 2008 {\em Adv. Phys.\/} {\bf 57} 225--285
  \urlprefix\url{https://doi.org/10.1080/00018730802218067}

\bibitem{IgnacioCirac2009}
Verstraete F, Wolf M and Ignacio~Cirac J 2009 {\em Nat. Phys.\/} {\bf 5} 633
  \urlprefix\url{https://www.nature.com/articles/nphys1342}

\bibitem{Devoret2013}
Shankar S {\em et~al.\/} 2013 {\em Nature\/} {\bf 504} 419
  \urlprefix\url{https://www.nature.com/articles/nature12802}

\bibitem{Kapit2017}
Kapit E 2017 {\em Quantum Sci. Technol.\/} {\bf 2} 033002
  \urlprefix\url{https://dx.doi.org/10.1088/2058-9565/aa7e5d}

\bibitem{Murch2022}
Harrington P~M, Mueller E~J and Murch K~W 2022 {\em Nat. Rev. Phys.\/} {\bf 4}
  660 \urlprefix\url{https://www.nature.com/articles/s42254-022-00494-8}

\bibitem{Rivas_2014}
Rivas A, Huelga S~F and Plenio M~B 2014 {\em Rep. Progr. Phys.\/} {\bf 77}
  094001 \urlprefix\url{https://dx.doi.org/10.1088/0034-4885/77/9/094001}

\bibitem{BreuerRMP2016}
Breuer H~P, Laine E~M, Piilo J and Vacchini B 2016 {\em Rev. Mod. Phys.\/} {\bf
  88}(2) 021002
  \urlprefix\url{https://link.aps.org/doi/10.1103/RevModPhys.88.021002}

\bibitem{Guerreschi2020}
Sawaya N~P~D, Menke T, Kyaw T, Johri S, Aspuru-Guzik A and Guerreschi G 2020
  {\em npj Quantum Info.\/} {\bf 6} 1
  \urlprefix\url{https://www.nature.com/articles/s41534-020-0278-0}

\bibitem{Zagoskin2021}
Andreev A~V, Balanov A~G, Fromhold T~M, Greenaway M~T, Hramov A~E, Li W,
  Makarov V~V and Zagoskin A~M 2021 {\em npj Quantum Info.\/} {\bf 7} 1
  \urlprefix\url{https://www.nature.com/articles/s41534-020-00339-1}

\bibitem{Lloyd96}
Lloyd S 1996 {\em Science\/} {\bf 273} 1073
  \urlprefix\url{https://www.science.org/doi/abs/10.1126/science.273.5278.1073}

\bibitem{Daley2014}
Daley A~J 2014 {\em Adv. Phys.\/} {\bf 63} 77
  \urlprefix\url{https://www.tandfonline.com/doi/full/10.1080/00018732.2014.933502}

\bibitem{NoriRMP2014}
Georgescu I~M, Ashhab S and Nori F 2014 {\em Rev. Mod. Phys.\/} {\bf 86}(1)
  153--185 \urlprefix\url{https://link.aps.org/doi/10.1103/RevModPhys.86.153}

\bibitem{DelCampo2017}
Chenu A, Beau M, Cao J and del Campo A 2017 {\em Phys. Rev. Lett.\/} {\bf 118}
  140403
  \urlprefix\url{https://journals.aps.org/prl/abstract/10.1103/PhysRevLett.118.140403}

\bibitem{NoriNPJ2018}
Wang B~X, Tao M~J, Ai Q, Xin T, Lambert N, Ruan D, Cheng Y~C, Nori F, Deng F~G
  and Long G~L 2018 {\em npj Quantum Info.\/} {\bf 4} 52
  \urlprefix\url{https://www.nature.com/articles/s41534-018-0102-2}

\bibitem{Daley2022}
Daley A~J, Bloch I, Kokail C, Flannigan S, Pearson N, Troyer M and Zoller P
  2022 {\em Nature\/} {\bf 607} 667
  \urlprefix\url{https://www.nature.com/articles/s41586-022-04940-6}

\bibitem{Jordan2022}
Kim C~W, Nichol J~M, Jordan A~N and Franco I 2022 {\em PRX Quantum\/} {\bf
  3}(4) 040308
  \urlprefix\url{https://link.aps.org/doi/10.1103/PRXQuantum.3.040308}

\bibitem{Kockum2025}
Chen G and Frisk~Kockum A 2025 {\em Quantum Sci. Technol.\/} {\bf 10} 025028
  \urlprefix\url{https://dx.doi.org/10.1088/2058-9565/adb2bd}

\bibitem{Chenu2024}
Martinez-Azcona P, Kundu A, Saxena A, del Campo A and Chenu A 2024 {\em arXiv
  preprint arXiv:2407.07746\/}

\bibitem{Burkard2023}
Gul\'acsi B and Burkard G 2023 {\em Phys. Rev. B\/} {\bf 107}(17) 174511
  \urlprefix\url{https://link.aps.org/doi/10.1103/PhysRevB.107.174511}

\bibitem{Petta2024}
Gullans M, Caranti M, Mills A and Petta J 2024 {\em PRX Quantum\/} {\bf 5}(1)
  010306 \urlprefix\url{https://link.aps.org/doi/10.1103/PRXQuantum.5.010306}

\bibitem{Gorini76}
Gorini V, Kossakowski A and Sudarshan E~C~G 1976 {\em J. Math. Phys.\/} {\bf
  17} 821--825 \urlprefix\url{https://doi.org/10.1063/1.522979}

\bibitem{Lindblad76}
Lindblad G 1976 {\em Commun. Math. Phys.\/} {\bf 48} 119
  \urlprefix\url{https://link.springer.com/article/10.1007/bf01608499}

\bibitem{Lloyd99}
Viola L, Knill E and Lloyd S 1999 {\em Phys. Rev. Lett.\/} {\bf 82}(12)
  2417--2421
  \urlprefix\url{https://link.aps.org/doi/10.1103/PhysRevLett.82.2417}

\bibitem{Zanardi99}
Zanardi P 1999 {\em Physics Letters A\/} {\bf 258} 77
  \urlprefix\url{https://www.sciencedirect.com/science/article/pii/S0375960199003655}

\bibitem{Vitali99}
Vitali D and Tombesi P 1999 {\em Phys. Rev. A\/} {\bf 59}(6) 4178
  \urlprefix\url{https://link.aps.org/doi/10.1103/PhysRevA.59.4178}

\bibitem{Lidar2005}
Khodjasteh K and Lidar D~A 2005 {\em Phys. Rev. Lett.\/} {\bf 95}(18) 180501
  \urlprefix\url{https://link.aps.org/doi/10.1103/PhysRevLett.95.180501}

\bibitem{DeLange2010}
De~Lange G, Wang Z, Riste D, Dobrovitski V and Hanson R 2010 {\em Science\/}
  {\bf 330} 60--63
  \urlprefix\url{https://www.science.org/doi/full/10.1126/science.1192739}

\bibitem{Coppersmith94}
L{\"o}fstedt R and Coppersmith S~N 1994 {\em Phys. Rev. Lett.\/} {\bf 72} 1947
  \urlprefix\url{https://journals.aps.org/prl/abstract/10.1103/PhysRevLett.72.1947}

\bibitem{Hanggi98}
Grifoni M and H{\"a}nggi P 1998 {\em Physics Reports\/} {\bf 304} 229
  \urlprefix\url{https://www.sciencedirect.com/science/article/pii/S0370157398000222}

\bibitem{Plenio2007}
Huelga S~F and Plenio M~B 2007 {\em Phys. Rev. Lett.\/} {\bf 98}(17) 170601
  \urlprefix\url{https://link.aps.org/doi/10.1103/PhysRevLett.98.170601}

\bibitem{Haug2019}
Wagner T, Talkner P, Bayer J~C, Rugeramigabo E~P, H{\"a}nggi P and Haug R~J
  2019 {\em Nature Physics\/} {\bf 15} 330
  \urlprefix\url{https://www.nature.com/articles/s41567-018-0412-5}

\bibitem{Kubo54}
Kubo R and Tomita K 1954 {\em J. Phys. Soc. Japan\/} {\bf 9} 888
  \urlprefix\url{https://doi.org/10.1143/JPSJ.9.888}

\bibitem{Bloch57}
Bloch F 1957 {\em Phys. Rev.\/} {\bf 105}(4) 1206--1222
  \urlprefix\url{https://link.aps.org/doi/10.1103/PhysRev.105.1206}

\bibitem{Redfield57}
Redfield A~G 1957 {\em IBM J. Res. Dev.\/} {\bf 1} 19
  \urlprefix\url{https://ieeexplore.ieee.org/abstract/document/5392713}

\bibitem{AbragamBK}
Abragam A 1986 {\em Principles of Nuclear Magnetism\/} (Oxford University Press
  (New York))

\bibitem{Mukamel78}
Mukamel S, Oppenheim I and Ross J 1978 {\em Phys. Rev. A\/} {\bf 17}(6) 1988
  \urlprefix\url{https://link.aps.org/doi/10.1103/PhysRevA.17.1988}

\bibitem{Kosloff95}
Geva E, Kosloff R and Skinner J~L 1995 {\em J. Chem. Phys.\/} {\bf 102}
  8541--8561 \urlprefix\url{https://doi.org/10.1063/1.468844}

\bibitem{Gulacsi2024}
Gul\'acsi B and Burkard G 2025 {\em Phys. Rev. Res.\/} {\bf 7}(2) 023073
  \urlprefix\url{https://link.aps.org/doi/10.1103/PhysRevResearch.7.023073}

\bibitem{Goldstein2019}
Shavit G, Horovitz B and Goldstein M 2019 {\em Phys. Rev. B\/} {\bf 100}(19)
  195436 \urlprefix\url{https://link.aps.org/doi/10.1103/PhysRevB.100.195436}

\bibitem{ClerkQuantum2023}
Groszkowski P, Seif A, Koch J and Clerk A~A 2023 {\em {Quantum}\/} {\bf 7} 972
  \urlprefix\url{https://doi.org/10.22331/q-2023-04-06-972}

\bibitem{SlichterBK}
Slichter C~P 1996 {\em Principles of Magnetic Resonance\/} (Springer-Verlag
  Berlin (Heidelberg))

\bibitem{PetruccioneBK}
Breuer H~P and Petruccione F 2002 {\em The theory of open quantum systems\/}
  (Oxford University Press (New York))

\bibitem{Geva2004}
Shi Q and Geva E 2004 {\em J. Chem. Phys.\/} {\bf 120} 10647
  \urlprefix\url{https://pubs.aip.org/aip/jcp/article/120/22/10647/534369/A-semiclassical-generalized-quantum-master}

\bibitem{Kubo66}
Kubo 1966 {\em Rep. Progr. Phys.\/} {\bf 29} 255
  \urlprefix\url{https://iopscience.iop.org/article/10.1088/0034-4885/29/1/306}

\bibitem{Redfield65}
Redfield A~G 1965 The theory of relaxation processes {\em Advances in Magnetic
  Resonance\/} vol~1 ed Waugh J~S (Academic Press) pp 1--32
  \urlprefix\url{https://www.sciencedirect.com/science/article/pii/B9781483231143500076}

\bibitem{ZwanzigBK}
Zwanzig R 2001 {\em Nonequilibrium statistical mechanics\/} (Oxford University
  Press (New York))

\bibitem{VanKampenBK}
{Van Kampen} N~G 2007 {\em Stochastic Processes in Physics and Chemistry\/}
  (North Holland (Amsterdam))

\bibitem{Petruccione99}
Breuer H~P, Kappler B and Petruccione F 1999 {\em Phys. Rev. A\/} {\bf 59}(2)
  1633 \urlprefix\url{https://link.aps.org/doi/10.1103/PhysRevA.59.1633}

\bibitem{Burgarth2017}
Burgarth D, Facchi P, Garnero G, Nakazato H, Pascazio S and Yuasa K 2017 {\em
  Op. Sys. Info. Dyn.\/} {\bf 24} 1750001 (\textit{Preprint}
  \eprint{https://doi.org/10.1142/S1230161217500019})
  \urlprefix\url{https://doi.org/10.1142/S1230161217500019}

\bibitem{AkhiezerBK}
Akhiezer N~I and Glazman I~M 2013 {\em Theory of linear operators in Hilbert
  space\/} (Frederick Ungar Publising co. (New York))

\bibitem{Fox75}
Fox R~F 1975 {\em J. Math. Phys.\/} {\bf 16} 289--297
  \urlprefix\url{https://aip.scitation.org/doi/abs/10.1063/1.522540}

\bibitem{Fox76}
Fox R~F 1976 {\em J. Math. Phys.\/} {\bf 17} 1148--1153
  \urlprefix\url{https://doi.org/10.1063/1.523041}

\bibitem{VanKampen74}
{Van Kampen} N~G 1974 {\em Physica\/} {\bf 74} 215
  \urlprefix\url{https://www.sciencedirect.com/science/article/pii/0031891474901219}

\bibitem{Terwiel74}
Terwiel R 1974 {\em Physica\/} {\bf 74} 248--265
  \urlprefix\url{https://www.sciencedirect.com/science/article/pii/0031891474901232}

\bibitem{Fox86}
Faid K and Fox R~F 1986 {\em Phys. Rev. A\/} {\bf 34} 4286
  \urlprefix\url{https://journals.aps.org/pra/abstract/10.1103/PhysRevA.34.4286}

\bibitem{Skinner87}
Budimir J and Skinner J 1987 {\em J. Stat. Phys.\/} {\bf 49} 1029
  \urlprefix\url{https://link.springer.com/article/10.1007/BF01017558}

\bibitem{Andersson2007}
Andersson E, Cresser J~D and Hall M~J~W 2007 {\em J. Mod. Opt.\/} {\bf 54} 1695
  \urlprefix\url{https://www.tandfonline.com/doi/full/10.1080/09500340701352581}

\bibitem{MukamelBK}
Mukamel S 1995 {\em Principles of Nonlinear Optical Spectroscopy\/} (Oxford
  University Press (New York))

\bibitem{Bernazzani2023}
Bernazzani L and Burkard G 2024 {\em Phys. Rev. Res.\/} {\bf 6}(1) 013284
  \urlprefix\url{https://link.aps.org/doi/10.1103/PhysRevResearch.6.013284}

\bibitem{Llobet2018}
Ko\l{}ody\ifmmode~\acute{n}\else \'{n}\fi{}ski J, Brask J~B, Perarnau-Llobet M
  and Bylicka B 2018 {\em Phys. Rev. A\/} {\bf 97}(6) 062124
  \urlprefix\url{https://link.aps.org/doi/10.1103/PhysRevA.97.062124}

\bibitem{Bloch53}
Wangsness R~K and Bloch F 1953 {\em Phys. Rev.\/} {\bf 89}(4) 728--739
  \urlprefix\url{https://link.aps.org/doi/10.1103/PhysRev.89.728}

\bibitem{Feynman57}
Feynman R~P, Vernon~Jr F~L and Hellwarth R~W 1957 {\em J. Appl. Phys.\/} {\bf
  28} 49
  \urlprefix\url{https://pubs.aip.org/aip/jap/article/28/1/49/161271/Geometrical-Representation-of-the-Schrodinger}

\bibitem{Burkard2004}
Burkard G, Koch R~H and DiVincenzo D~P 2004 {\em Phys. Rev. B\/} {\bf 69}(6)
  064503 \urlprefix\url{https://link.aps.org/doi/10.1103/PhysRevB.69.064503}

\bibitem{Lidar2007}
Krovi H, Oreshkov O, Ryazanov M and Lidar D~A 2007 {\em Phys. Rev. A\/} {\bf
  76}(5) 052117
  \urlprefix\url{https://link.aps.org/doi/10.1103/PhysRevA.76.052117}

\bibitem{Chaturvedi79}
Chaturvedi S and Shibata F 1979 {\em Z. Physik B\/} {\bf 35} 297
  \urlprefix\url{https://link.springer.com/article/10.1007/BF01319852}

\bibitem{Lidar2020}
Mozgunov E and Lidar D~A 2020 {\em {Quantum}\/} {\bf 4} 227 ISSN 2521
  \urlprefix\url{https://doi.org/10.22331/q-2020-02-06-227}

\bibitem{Uhlenbeck30}
Uhlenbeck G~E and Ornstein L~S 1930 {\em Phys. Rev.\/} {\bf 36}(5) 823
  \urlprefix\url{https://link.aps.org/doi/10.1103/PhysRev.36.823}

\bibitem{Novikov65}
Novikov E~A 1965 {\em Sov. Phys. JETP\/} {\bf 20} 1290

\bibitem{Budini2000}
Budini A~A 2000 {\em Phys. Rev. A\/} {\bf 63}(1) 012106
  \urlprefix\url{https://link.aps.org/doi/10.1103/PhysRevA.63.012106}

\bibitem{Budini2001}
Budini A~A 2001 {\em Phys. Rev. A\/} {\bf 64}(5) 052110
  \urlprefix\url{https://link.aps.org/doi/10.1103/PhysRevA.64.052110}

\bibitem{Kiely2021}
Kiely A 2021 {\em Europhysics Letters\/} {\bf 134} 10001
  \urlprefix\url{https://iopscience.iop.org/article/10.1209/0295-5075/134/10001/meta}

\bibitem{Garr1}
Garraway B~M 1997 {\em Phys. Rev. A\/} {\bf 55}(3) 2290--2303
  \urlprefix\url{https://link.aps.org/doi/10.1103/PhysRevA.55.2290}

\bibitem{Garr2}
Mazzola L, Maniscalco S, Piilo J, Suominen K~A and Garraway B~M 2009 {\em Phys.
  Rev. A\/} {\bf 80}(1) 012104
  \urlprefix\url{https://link.aps.org/doi/10.1103/PhysRevA.80.012104}

\bibitem{Garr3}
Pleasance G, Garraway B~M and Petruccione F 2020 {\em Phys. Rev. Res.\/} {\bf
  2}(4) 043058
  \urlprefix\url{https://link.aps.org/doi/10.1103/PhysRevResearch.2.043058}

\bibitem{Damanet2024}
Debecker B, Martin J and Damanet F~m~c 2024 {\em Phys. Rev. Lett.\/} {\bf
  133}(14) 140403
  \urlprefix\url{https://link.aps.org/doi/10.1103/PhysRevLett.133.140403}

\bibitem{Eisfeld2014}
Ritschel G and Eisfeld A 2014 {\em The Journal of chemical physics\/} {\bf 141}

\bibitem{Plenio2018}
Tamascelli D, Smirne A, Huelga S~F and Plenio M~B 2018 {\em Phys. Rev. Lett.\/}
  {\bf 120}(3) 030402
  \urlprefix\url{https://link.aps.org/doi/10.1103/PhysRevLett.120.030402}

\bibitem{Lambert2019}
Lambert N, Ahmed S, Cirio M and Nori F 2019 {\em Nat. Commun.\/} {\bf 10} 3721
  \urlprefix\url{https://doi.org/10.1038/s41467-019-11656-1}

\bibitem{Stockburger2022}
Xu M, Yan Y, Shi Q, Ankerhold J and Stockburger J~T 2022 {\em Phys. Rev.
  Lett.\/} {\bf 129}(23) 230601
  \urlprefix\url{https://link.aps.org/doi/10.1103/PhysRevLett.129.230601}

\bibitem{Hartmann2020}
Hartmann R and Strunz W~T 2020 {\em Phys. Rev. A\/} {\bf 101}(1) 012103
  \urlprefix\url{https://link.aps.org/doi/10.1103/PhysRevA.101.012103}

\bibitem{DiBari92}
Levitt M~H and Di~Bari L 1992 {\em Phys. Rev. Lett.\/} {\bf 69}(21) 3124
  \urlprefix\url{https://link.aps.org/doi/10.1103/PhysRevLett.69.3124}

\bibitem{Ernst95}
Levante T and Ernst R 1995 {\em Chem. Phys. Lett.\/} {\bf 241} 73

\bibitem{Scarlino2025}
Peyruchat L, Minganti F, Scigliuzzo M, Ferrari F, Jouanny V, Nori F, Savona V
  and Scarlino P 2025 {\em npj Quantum Info.\/} {\bf 11} 62
  \urlprefix\url{https://www.nature.com/articles/s41534-025-00984-4}

\bibitem{Lidar2025}
Saurav K and Lidar D~A 2025 {\em PRX Quantum\/} {\bf 6}(1) 010335
  \urlprefix\url{https://link.aps.org/doi/10.1103/PRXQuantum.6.010335}

\bibitem{Andersen2025}
Zwanenburg M~F, Singh S, Huang E~Y, Yilmaz F, Stefanski T~V, Hu J,
  Kumaravadivel P and Andersen C~K 2025 {\em arXiv:2503.08238\/}

\bibitem{Gasparinetti2025}
Yang J, Strandberg I, Vivas-Via{\~n}a A, Gaikwad A, Castillo-Moreno C, Kockum
  A~F, Ullah M~A, Mu{\~n}oz C~S, Eriksson A~M and Gasparinetti S 2025 {\em npj
  Quantum Info.\/} {\bf 11} 1
  \urlprefix\url{https://www.nature.com/articles/s41534-025-00995-1}

\bibitem{Grifoni2000}
Hartmann L, Goychuk I, Grifoni M and H\"anggi P 2000 {\em Phys. Rev. E\/} {\bf
  54}(2) R4687
  \urlprefix\url{https://journals.aps.org/pre/abstract/10.1103/PhysRevE.61.R4687}

\bibitem{Flindt2009}
Flindt C, Fricke C, Hohls F, Novotn{\`y} T, Neto{\v{c}}n{\`y} K, Brandes T and
  Haug R~J 2009 {\em Proc. Natl. Acad. Sci. USA\/} {\bf 106} 10116

\bibitem{Haug2025}
Bayer J~C, Brange F, Schmidt A, Wagner T, Rugeramigabo E~P, Flindt C and Haug
  R~J 2025 {\em Phys. Rev. Lett.\/} {\bf 134}(4) 046303
  \urlprefix\url{https://link.aps.org/doi/10.1103/PhysRevLett.134.046303}

\bibitem{Mollow69}
Mollow B~R 1969 {\em Phys. Rev.\/} {\bf 188}(5) 1969
  \urlprefix\url{https://link.aps.org/doi/10.1103/PhysRev.188.1969}

\bibitem{Mork2012}
Moelbjerg A, Kaer P, Lorke M and M\o{}rk J 2012 {\em Phys. Rev. Lett.\/} {\bf
  108}(1) 017401
  \urlprefix\url{https://link.aps.org/doi/10.1103/PhysRevLett.108.017401}

\bibitem{Hughes2018}
Gustin C, Manson R and Hughes S 2018 {\em Opt. Lett.\/} {\bf 43} 779
  \urlprefix\url{https://opg.optica.org/ol/fulltext.cfm?uri=ol-43-4-779&id=381516}

\bibitem{Mueller2024}
Boos K, Kim S~K, Bracht T, Sbresny F, Kaspari J~M, Cygorek M, Riedl H, Bopp
  F~W, Rauhaus W, Calcagno C, Finley J~J, Reiter D~E and M\"uller K 2024 {\em
  Phys. Rev. Lett.\/} {\bf 132}(5) 053602
  \urlprefix\url{https://link.aps.org/doi/10.1103/PhysRevLett.132.053602}

\bibitem{Stobinska2024}
L{\'o}pez~Carre{\~n}o J~C, Berm{\'u}dez~Feijoo S and Stobi{\'n}ska M 2024 {\em
  npj Nanophoton.\/} {\bf 1} 3

\bibitem{Breuer2007}
Breuer H~P 2007 {\em Phys. Rev. A\/} {\bf 75}(2) 022103
  \urlprefix\url{https://link.aps.org/doi/10.1103/PhysRevA.75.022103}

\end{thebibliography}

\end{document}